\documentclass[onecolumn,aps,prd,amsmath,amssymb]{revtex4-1}
\usepackage[latin9]{inputenc}
\usepackage{amstext}
\usepackage{amssymb}
\usepackage{amsmath}
\usepackage{graphicx}
\usepackage{esint}

\makeatletter
\usepackage{amssymb}
\usepackage{mathbbold}
\usepackage{amsfonts}
\usepackage{epsfig}
\usepackage{bm}\usepackage{dcolumn}\usepackage{color}
\usepackage{overpic}
\usepackage[bottom]{footmisc}

\makeatother

\begin{document}

\title{Response functions of hot and dense matter in the Nambu-Jona-Lasino model}

\author{Chengfu Mu$^{1}$}

\author{Ziyue Wang$^{2}$}

\author{Lianyi He$^{2,3,4}$}

\affiliation{$^{1}$ School of Science, Huzhou University, Zhejiang 313000, China}

\affiliation{$^{2}$ Department of Physics, Tsinghua University, Beijing 100084, China}

\affiliation{$^{3}$ State Key Laboratory of Low-Dimensional Quantum Physics, Tsinghua University, Beijing 100084, China}

\affiliation{$^{4}$ Collaborative Innovation Center of Quantum Matter, Beijing 100084, China}

\date{\today}

\begin{abstract}
We investigate the current-current correlation functions or the so-called response functions of a two-flavor Nambu-Jona-Lasino model at finite temperature and density.  We study the linear response by using the functional path integral approach and introducing the conjugated gauge fields as external sources. The response functions can be obtained by expanding the generational functional in powers of the external sources.  We derive the response functions parallel to two well-established approximations for the equilibrium thermodynamics: the mean-field theory and a beyond-mean-field theory taking into account the mesonic contributions. The response functions based on the mean-field theory recover the so-called quasiparticle random phase approximation. We calculate the dynamical structure factors for the density responses in various channels within the random phase approximation.  We show that the dynamical structure factors in the baryon axial vector and isospin axial vector channels can be used reveal the quark mass gap and the Mott dissociation of mesons, respectively. Noting that the mesonic contributions are not taken into account in the random phase approximation, we also derive the response functions parallel to the beyond-mean-field theory.  We show that the mesonic fluctuations naturally give rise to three kinds of famous diagrammatic contributions: the Aslamazov-Lakin contribution, the Self-Energy or Density-of-State contribution, and the Maki-Thompson contribution. Unlike the equilibrium case, in evaluating the fluctuation contributions, we need to treat carefully the linear terms in the external sources and the induced perturbations.  In the chiral symmetry breaking phase, we find an additional chiral order parameter induced contribution, which ensures that the temporal component of the response functions in the static and long-wavelength limit recovers the correct charge susceptibility defined by using the equilibrium thermodynamic quantities. These contributions from the mesonic fluctuations are expected to have significant effects on the transport properties of hot and dense matter around the chiral phase transition or crossover, where the mesonic degrees of freedom are still important.

\end{abstract}

\maketitle

\section{Introduction}
A good knowledge of strongly interacting matter, i.e., quantum chromodynamics (QCD) at nonzero temperature and density, is important for us to understand many
physical phenomena in nature. For instance, the nature of the QCD phase transition at temperature around $200$MeV and at vanishingly small baryon density \cite{Kolb2004,Rischke2004} is needed for us to understand the evolution of the early universe. On the other hand, the nature of high-density QCD matter at very low temperature \cite{Rischke2004,Alford2008,Buballa2005,Shovkovy2005,Huang2005,Wang2010,Fukushima2011,Anglani2014}is crucial for us to explain the phenomenology of neutron stars.  It has been shown that QCD has a very rich phase structure at high baryon density due to the appearance of color superconductivity \cite{Rischke2004,Alford2008,Buballa2005,Shovkovy2005,Huang2005,Wang2010,Fukushima2011,Anglani2014}. 

At ultra high temperature and/or baryon density, perturbative method can be applied to predict the phases and equation of state of hot and dense QCD matter \cite{Kapusta-Book,PQCD-T,PQCD-MU}. However, near the QCD phase transition, the
system is strongly interacting and hence the usual perturbative method fails. One powerful nonperturbative method, the lattice simulation of QCD at nonzero temperature and vanishing baryon density, has reached great success in the past decades \cite{Borsanyi2010a,Borsanyi2010b,Bazavov2012,Bazavov2014}.  However, at nonzero baryon density there exists the so-called sign problem \cite{Karsch2002,Muroya2003}: The fermion determinant is not generally a complex number and hence cannot be regarded as probability.  Therefore, no satisfying lattice results at nonzero baryon density have been achieved so far. Another useful nonperturbative method is the functional renormalization group \cite{Wetterich1993,Berges2002}, which has made great progress in understanding the QCD phase transitions \cite{Berges1999,Braun2011,Mitter2015}.
 
While QCD itself is hard to handle, it is generally believed that a number of features of QCD phase transitions can be captured by some low-energy effective models of QCD. One of these effective models, the
Nambu--Jona-Lasinio (NJL) model \cite{Nambu1961}, with quarks as elementary degrees of freedom, can describe well the low-energy phenomenology of the QCD vacuum \cite{Volkov1984,Weise1991,Klevansky1992,Hatsuda1994}. It is generally believed that the NJL model still works well at low and moderate temperature and density \cite{Klevansky1992,Hatsuda1994}.  One disadvantage of this model, i.e., the lack of confinement of quarks, has been  amended by the so-called Polyakov loop extended NJL model \cite{Fukushima2004,Ratti2006,Roessner2007,Fukushima2008,Sasaki2007,Ghosh2006,Abuki2008,Fu2008}. As a pure fermionic field theoretical model with contact four-fermion interactions, some nonperturbative method from condensed matter theory can be applied. One simple but useful approximation is the mean-field theory, which gives a reasonable description of the chiral phase transition. The mesons can be constructed by using the random phase approximation \cite{Klevansky1992,Hatsuda1994}.  However, because of the strong coupling nature, the mean-field theory is not adequate: (1) The thermodynamic quantities lacks the mesonic degrees of freedom in the chiral symmetry breaking phase, or the hadronic phase at low temperature, where it is believed that the pions dominate the thermodynamical quantities; 
(2) In the chiral limit, the quarks become massless above the chiral phase transition temperature, and hence the mean-field theory predicts a gas of noninteracting massless quarks. These inadequacies indicates that
going beyond mean field, i.e., taking into account properly the mesonic degrees of freedom, is quite necessary both below and above the chiral phase transition temperature.

Such a system is very similar to the BCS-BEC crossover in strongly interacting Fermi gases \cite{BCSBEC01,BCSBEC02,BCSBEC03,BCSBEC04,BCSBEC05,BCSBEC06,BCSBEC07,BCSBEC08,BCSBEC09}. There, it has been shown that the role of pair degree of freedom is of significant important to describe quantitatively the equation
of state and other properties of the BCS-BEC crossover \cite{PAIR01,PAIR02,PAIR03,PAIR04,PAIR05,PAIR06,PAIR07,PAIR08,PAIR09,PAIR10,PAIR11,PAIR12,PAIR13,PAIR14,PAIR15}. The Gaussian approximation for the pair fluctuations, which truncates the pair fluctuations at the two-body level, has achieved great success in describing quantitatively the the equation
of state in  BCS-BEC crossover, both in three and two spatial dimensions \cite{PAIR10,PAIR11,PAIR12,PAIR13,PAIR14,PAIR15}.  For the NJL model,   the parallel Gaussian approximation which includes the mesonic degrees of freedom has been developed by  Huefner,  Klevansky, Zhuang, and Voss \cite{Hufner1994}.
At low temperature, such a beyond-mean-field theory predicts that the thermodynamical quantities are dominated by the lightest mesonic excitations, i.e., the pions  \cite{Zhuang1994}. On the other hand, it has been shown that the mesonic fluctuations or the fluctuations of the chiral order parameter are also important above and near the chiral phase transition temperature \cite{Hatsuda1985,Kitazawa2006}. In dense quark matter, the corresponding diquark fluctuation is expected to give significant contribution to the transport properties above and near the transition temperature of color superconductivity \cite{Kitazawa2002,Kerbikov01,Kerbikov02}.

In this work, we derive the current-current correlation functions or the so-called response functions of a two-flavor Nambu-Jona-Lasino model at finite temperature and density. We study the linear response by using the functional path integral approach and introducing the conjugated gauge fields as external sources. The response functions can be obtained by expanding the generational functional in powers of the external sources \cite{He2016}. We will derive the response functions parallel to two well-established approximations for the equilibrium thermodynamics: the mean-field theory \cite{Klevansky1992,Hatsuda1994} and a beyond-mean-field theory taking into account the mesonic contributions \cite{Hufner1994,Zhuang1994}.  The latter beyond-mean-field theory can be called the meson-fluctuation theory. The response functions based on the mean-field theory recover the so-called quasiparticle random phase approximation. The dynamical structure factors for various density responses are evaluated. It has been shown that in the long-wavelength limit, the dynamical structure factor is nonzero only for the baryon axial vector and isospin axial vector channels. For the isospin axial vector channel, the dynamical density response couples to the pion, and hence the corresponding dynamical structure factor can be used to reveal the Mott dissociation of mesons at finite temperature \cite{Hatsuda1994,Blaschke2013,Blaschke2017}. Below the Mott transition temperature, the dynamical structure factor reveals a pole plus continuum structure. Above the Mott transition temperature, the dynamical structure factor displays only a continuum. 

We find that the random phase approximation becomes inadequate above the chiral phase transition temperature: In the chiral limit it describes the linear response of a hot gas of noninteracting massless quarks. We thus 
further develop a linear response theory parallel to the meson-fluctuation theory which includes properly the mesonic degrees of freedom.  We show that the mesonic fluctuations naturally give rise to three kinds of famous diagrammatic contributions: the Aslamazov-Lakin contribution \cite{AL1968}, the Self-Energy or Density-of-State contribution, and the Maki-Thompson contribution \cite{MT1968}. Unlike the equilibrium case, in evaluating the fluctuation contributions, we need to treat carefully the linear terms in the external sources and the induced order parameter perturbations.  In the chiral symmetry breaking phase, we find an additional chiral order parameter induced contribution, which ensures that the temporal component of the response functions in the static and long-wavelength limit recovers the correct charge susceptibility defined by using the equilibrium thermodynamic quantities. These contributions from the mesonic fluctuations are expected to have significant effects on the transport properties of hot and dense matter around the chiral phase transition or crossover, where the mesonic degrees of freedom are still important.

We organize this paper as follows. In Sec. \ref{s2}, we review the two-flavor NJL model and its vacuum phenomenology. In Sec. \ref{s3}, we review the thermodynamics of the NJL model in the mean-field theory and the meson-fluctuation theory by using the path integral approach. In Sec. \ref{s4}, we introduce the general linear response theory for the current-current correlations in the path integral approach. In Sec. \ref{s5}, we evaluate the response functions in the mean-field theory, which recovers the quasiparticle random phase approximation from the diagrammatic point of view.  In Sec. \ref{s6}, we evaluate the dynamical structure factors for the density responses in various channels. In Sec. \ref{s7}, we consider the role of meson fluctuations and develop a linear response theory for the NJL model beyond the random phase approximation. We summarize in Sec. \ref{s8}. We use the natural units $c=\hbar=k_{\rm B}=1$ throughout.

\section{Nambu--Jona-Lasino model}\label{s2}

For a general $N_f$-flavor Nambu--Jona-Lasinio model, the Lagrangian density is given by \cite{Klevansky1992}
\begin{eqnarray}
{\cal L}_{\rm{NJL}}&=&\bar{\psi}(i\gamma^\mu\partial_\mu-\hat{m}_{\rm c})\psi+{\cal
L}_{\rm S}+{\cal L}_{\rm{KMT}},\nonumber\\
{\cal L}_{\rm S}&=&G_{\rm
s}\sum_{\alpha=0}^{N_f^2-1}\left[\left(\bar{\psi}\lambda_\alpha
\psi\right)^2+\left(\bar{\psi}i\gamma_5\lambda_\alpha\psi\right)^2\right],\nonumber\\
{\cal
L}_{\rm{KMT}}&=&-K\left[\det\bar{\psi}\left(1+\gamma_5\right)\psi+\det\bar{\psi}\left(1-\gamma_5\right)\psi\right],
\end{eqnarray}
where $\lambda_\alpha$ $(\alpha=0,1,\cdots,N_f^2-1)$ is the $N_f$-flavor Gell-Mann matrix with $\lambda_0=\sqrt{2/N_f}$ and   
$\hat{m}_{\rm c}={\rm diag}(m_{\rm u},m_{\rm d},m_{\rm s},\cdots)$ is the current quark mass matrix. In the special case $m_{\rm u}=m_{\rm d}=m_{\rm s}=\cdots=0$ and $K=0$, ${\cal L}_{\rm{
NJL}}$ is invariant under the group transformation ${\rm SU_C}(N_c)\otimes {\rm SU_V}(N_f)\otimes{\rm SU_A}(N_f)\otimes \rm{U}_B(1)\otimes
\rm{U}_A(1)$. ${\cal L}_{\rm{KMT}}$ is the so-called Kobayashi-Maskawa-t'Hooft term with $K<0$ is designed to break the ${\rm U}_{\rm A}(1)$ symmetry.
For the three-flavor case ($N_f=3$), ${\cal L}_{\rm{KMT}}$ contains six-fermion interactions and can describe well the mass splitting between $\eta$ and $\eta^\prime$. In this work, we consider
the two-flavor case, where ${\cal L}_{\rm{KMT}}$ contains only four-fermion interactions, like the mesonic interaction term ${\cal L}_{\rm S}$. The Lagrangian density of the general two-flavor NJL model 
is given by
\begin{eqnarray}
{\cal L}_{\rm{ NJL}}=\bar\psi(i\gamma^\mu\partial_\mu-m_0)\psi
+G\left[(\bar{\psi}\psi)^{2}+(\bar{\psi}i\gamma_{5}\mbox{\boldmath{$\tau$}}\psi)^{2}\right]
+G^\prime\left[(\bar{\psi}\mbox{\boldmath{$\tau$}}\psi)^{2}+(\bar{\psi}i\gamma_{5}\psi)^{2}\right],
\end{eqnarray}
where $G=G_{\rm s}-K, G^\prime=G_{\rm s}+K$, and we assume $m_{\rm u}=m_{\rm d}=m_0$. Since the masses of  scalar-isovector and pseudoscalcar-isoscalar mesons in the two-flavor case are much larger than the sigma meson and pions, we consider the maximal axial symmetry breaking case $|K|=G_{\rm s}$, which leads to the minimal NJL model
\begin{eqnarray}
{\cal L}_{\rm{ NJL}}=\bar\psi(i\gamma^\mu\partial_\mu-m_0)\psi
+G\left[(\bar{\psi}\psi)^{2}+(\bar{\psi}i\gamma_{5}\mbox{\boldmath{$\tau$}}\psi)^{2}\right].
\end{eqnarray}
In this work, we study this minimum NJL model for the sake of simplicity.

In the functional path integral formalism, the partition function of the NJL model can
be written as
\begin{eqnarray}
{\cal Z}_{\rm NJL}=\int [d\psi][d\bar{\psi}] \exp\left\{ i\int d^4x\cal{L}_{\rm
NJL}\right\}.
\end{eqnarray}
Introduce two auxiliary fields $\sigma$ and $\mbox{\boldmath{$\pi$}}$ which satisfy
equations of motion $\sigma=-2G\bar{\psi}\psi,\mbox{\boldmath{$\pi$}}=-2G\bar{\psi}
i\gamma_5\mbox{\boldmath{$\tau$}}\psi$, and apply the Hubbard-Strotonovich
transformation, we obtain
\begin{eqnarray}
{\cal Z}_{\rm NJL}=\int [d\psi] [d\bar{\psi}] [d\sigma]
[d\mbox{\boldmath{$\pi$}}]\exp\bigg\{i{\cal
S}[\psi,\bar{\psi},\sigma,\mbox{\boldmath{$\pi$}}]\bigg\},
\end{eqnarray}
where the action reads
\begin{eqnarray}
{\cal S}[\psi,\bar{\psi},\sigma,\mbox{\boldmath{$\pi$}}]&=&-\int d^4x\frac{\sigma^2+\mbox{\boldmath{$\pi$}}^2}{4G}+\int d^4x\int d^4x^\prime\bar{\psi}(x){\bf G}^{-1}(x,x^\prime)\psi(x^\prime),\nonumber\\
{\bf
G}(x,x^\prime)&=&\left[i\gamma^\mu\partial_\mu-m_0-(\sigma+i\gamma_5\mbox{\boldmath{$\tau$}}\cdot\mbox{\boldmath{$\pi$}})\right]\delta(x-x^\prime).
\end{eqnarray}
Then we integrate out the quark field and obtain
\begin{eqnarray}
{\cal Z}_{\rm NJL}&=&\int [d\sigma][d\mbox{\boldmath{$\pi$}}] \exp\left\{i{\cal
S}_{\rm{eff}}[\sigma,\mbox{\boldmath{$\pi$}}]\right\},\nonumber\\
{\cal S}_{\rm{eff}}[\sigma,\mbox{\boldmath{$\pi$}}]&=&-\frac{1}{4G}\int
d^4x(\sigma^2+\mbox{\boldmath{$\pi$}}^2)-i{\rm Tr}\ln{\bf G}^{-1}(x,x^\prime).
\end{eqnarray}

The partition function cannot be evaluated precisely. We assume that the sigma field
acquires a nonvanishing expectation value $\langle\sigma(x)\rangle=\upsilon$ and set
$\langle\mbox{\boldmath{$\pi$}}(x)\rangle={\bf 0}$, which characterizes the dynamical
chiral symmetry breaking (DCSB). Then the auxiliary fields can be expanded around their
expectation values. After making the field shifts, $\sigma(x)\rightarrow\upsilon+\sigma(x)$ and
$\mbox{\boldmath{$\pi$}}(x)\rightarrow{\bf 0}+\mbox{\boldmath{$\pi$}}(x)$, we expand the effective
action ${\cal S}_{\rm{eff}}[\sigma,\mbox{\boldmath{$\pi$}}]$ in powers
of the fluctuations $\sigma(x)$ and $\mbox{\boldmath{$\pi$}}(x)$. We have
\begin{equation}
{\cal S}_{\rm{eff}}[\sigma,\mbox{\boldmath{$\pi$}}]={\cal S}_{\rm{eff}}^{(0)}+{\cal
S}_{\rm{eff}}^{(1)}[\sigma,\mbox{\boldmath{$\pi$}}] +{\cal
S}_{\rm{eff}}^{(2)}[\sigma,\mbox{\boldmath{$\pi$}}]+\cdots.
\end{equation}
The mean-field part ${\cal S}_{\rm{eff}}^{(0)}={\cal S}_{\rm{eff}}[\upsilon,{\bf 0}]$ can be evaluated as
\begin{eqnarray}\label{1-45}
\frac{{\cal S}_{\rm{eff}}^{(0)}}{V_4}=\frac{\upsilon^2}{4G}-2N_cN_f\int\frac{d^3{\bf
k}}{(2\pi)^3}E_{\bf k},
\end{eqnarray}
where $E_{\bf k}=\sqrt{{\bf k}^2+M^2}$ with the effective quark mass $M=m_0+\upsilon$.  Since the NJL model is not renormalizable, we employ a hard cutoff $\Lambda$ to regularize the integral over the quark momentum 
${\bf k}$ ($|{\bf k}|<\Lambda$). The condensate $\upsilon$
should be determined by minimizing ${\cal S}_{\rm{eff}}^{(0)}$, i.e., $\partial {\cal
S}_{\rm{eff}}^{(0)}/\partial\upsilon=0$, which gives rise to the gap equation
\begin{eqnarray}\label{NJL-gap0}
M-m_0=4GN_cN_fM\int\frac{d^3{\bf k}}{(2\pi)^3}\frac{1}{E_{\bf k}}.
\end{eqnarray}
In the chiral limit $m_0=0$, we find that if $G>\pi^2/(N_cN_f\Lambda^2)$ \cite{Klevansky1992,Hatsuda1994}, the sigma field acquires a nonvanishing expectation value
$\upsilon\neq0$ and hence the DCSB occurs.

The gap equation (\ref{NJL-gap0}) ensures that the linear term ${\cal S}_{\rm{eff}}^{(1)}[\sigma,\mbox{\boldmath{$\pi$}}]$ vanishes. The mesons in the NJL
model are regarded as collective excitations, which are characterized by the Gaussian
fluctuation term ${\cal S}_{\rm{eff}}^{(2)}[\sigma,\mbox{\boldmath{$\pi$}}]$. Using the
derivative expansion
\begin{eqnarray}
{\rm Tr}\ln{(1-{\cal G}\Sigma)}=-\sum_{n=1}^\infty\frac{1}{n}{\rm Tr}({\cal G}\Sigma)^n,
\end{eqnarray}
with ${\cal G}=(\gamma^\mu K_\mu-M)^{-1}$ being the mean-field quark propagator and
$\Sigma=\sigma+i\gamma_5\mbox{\boldmath{$\tau$}}\cdot\mbox{\boldmath{$\pi$}}$, we obtain
\begin{eqnarray}\label{1-48}
 {\cal S}_{\rm eff}^{(2)}[\sigma,\mbox{\boldmath{$\pi$}}]&=&-\frac{1}{2}\int\frac{d^4Q}{(2\pi)^4}
\left[{\cal D}_\sigma^{-1}(Q)\sigma(Q)\sigma(-Q)+{\cal D}_{\pi}^{-1}(Q)\mbox{\boldmath{$\pi$}}(Q)\cdot\mbox{\boldmath{$\pi$}}(-Q)\right],\nonumber\\
{\cal D}^{-1}_{\sigma,\pi}(Q)&=&\frac{1}{2G}-\Pi_{\sigma,\pi}(Q).
\end{eqnarray}
Here the polarization functions $\Pi_{\sigma,\pi}(Q)$ are given by
\begin{eqnarray}
\Pi_{\sigma,\pi}(Q)&=&4iN_cN_f\int\frac{d^4K}{(2\pi)^4}\frac{1}{K^2-M^2}-2iN_cN_f(Q^2-\varepsilon_{\sigma,\pi}^2)I(Q^2),\nonumber\\
I(Q^2)&=&\int\frac{d^4K}{(2\pi)^4}\frac{1}{[(K+Q/2)^2-M^2][(K-Q/2)^2-M^2]},
\end{eqnarray}
with $\varepsilon_\sigma=2M$ and $\varepsilon_\pi=0$.

The masses of the mesons are determined by the pole of their propagators,
i.e., ${\cal D}^{-1}_{\sigma,\pi}(Q^2=m_{\sigma,\pi}^2)=0$. We obtain
\begin{eqnarray}
m_{\sigma,\pi}^2=-\frac{m_0}{M}\frac{1}{4iGN_cN_fI(m_{\sigma,\pi}^2)}+\varepsilon_{\sigma,\pi}^2.
\end{eqnarray}
Note that the function $I(Q^2)$ changes very slowly with $Q^2$. Therefore, we can
approximate $I(m_{\sigma,\pi}^2)\approx I(0)$.  The meson masses are given by
\begin{eqnarray}
m_{\pi}^2\approx-\frac{m_0}{M}\frac{1}{4iGN_cN_fI(0)},\ \ \ \ \ \ m_\sigma^2\approx
m_\pi^2+4M^2.
\end{eqnarray}
Near the poles, the meson propagators can be well approximated as
\begin{eqnarray}
{\cal D}_{\sigma,\pi}(Q)\simeq\frac{g^2_{\sigma qq,\pi qq}}{Q^2-m_{\sigma,\pi}^2},
\end{eqnarray}
where the meson-quark couplings are given by
\begin{eqnarray}\label{gmqq}
g^{-2}_{\sigma qq,\pi qq}\equiv \frac{\partial\Pi_{\sigma,\pi}}{\partial
Q^2}\bigg|_{Q^2=m_{\sigma,\pi}^2}\approx-2iN_cN_fI(0).
\end{eqnarray}

To determine the model parameters, i.e., the current quark mass $m_0$, the coupling
constant $G$, and the cutoff $\Lambda$, we need to derive the pion decay constant
$f_\pi$ in the NJL model. It can be obtained by calculating the matrix element of the vacuum to one-pion
axial-vector current transition. We have
\begin{eqnarray}
iQ_\mu f_\pi\delta^{ij}&=&-{\rm Tr}\int\frac{d^4K}{(2\pi)^4}\left[ i\gamma_\mu
\gamma_5\frac{\tau^i}{2}i{\cal G}(K+Q/2)ig_{\pi qq}\gamma_5\tau^ji{\cal
G}(K-Q/2)\right]\nonumber\\
&=&2N_cN_fg_{\pi qq}M Q_\mu I(Q^2)\delta^{ij}.
\end{eqnarray}
Using Eq. (\ref{gmqq}), we obtain
\begin{eqnarray}
f_\pi^2\approx-2iN_cN_fM^{2}I(0).
\end{eqnarray}
Applying the result $M=-2G\langle\bar{\psi}\psi\rangle_0+m_0$, we recover the
Gell-Mann-Oakes-Renner relation
\begin{eqnarray}
m_\pi^2f_\pi^2\approx-m_0\langle\bar{\psi}\psi\rangle_0.
\end{eqnarray}
The model parameters can be fixed by matching the pion mass $m_\pi$, the pion decay
constant $f_\pi$, and the chiral condensate $\langle\bar{\psi}\psi\rangle_0$.  For the
physical case, we choose $m_0=5{\rm MeV}$, $G=4.93{\rm GeV}^{-2}$, and $\Lambda=653{\rm
MeV}$, which gives $m_\pi=134{\rm MeV}$, $f_\pi=93{\rm MeV}$, and
$\langle\bar{u}u\rangle_0=-(250{\rm MeV})^3$. In the chiral limit, $m_0=0$, we use
$G=5.01{\rm GeV}^{-2}$, and $\Lambda=650{\rm MeV}$.

\section{Phase Diagram and Thermodynamics of the NJL model}\label{s3}
The partition function of the NJL model at finite temperature $T$ can be given by the
imaginary time formalism,
\begin{eqnarray}
{\cal Z}_{\rm NJL} = \int [d\psi][d\bar{\psi}]\exp\left\{\int dx \left[{\cal L}_{\rm
NJL}+\bar\psi\hat{\mu}\gamma^0\psi\right]\right\}.
\end{eqnarray}
Here and in the following, $x=(\tau,{\bf r})$ with $\tau$ being the imaginary time. We use the notation $\int dx\equiv \int_0^\beta d\tau\int d^3{\bf r}$
with $\beta=1/T$. The chemical potential matrix $\hat{\mu}$ is diagonal in flavor space,
$\hat{\mu}={\rm diag}(\mu_{\rm u},\mu_{\rm d})$. A useful parameterization of the
chemical potentials is given by
\begin{eqnarray}
\mu_{\rm u}=\frac{1}{3}\mu_{\rm B}+\frac{1}{2}\mu_{\rm I},\ \ \ \ \ \ \ \ \ \mu_{\rm
d}=\frac{1}{3}\mu_{\rm B}-\frac{1}{3}\mu_{\rm I},
\end{eqnarray}
corresponding to introducing two conserved charges, the baryon number and the third
component of the isospin. In this work, we consider the case $\mu_{\rm I}=0$ for the
sake of simplicity. We therefore set $\mu_{\rm u}=\mu_{\rm d}\equiv\mu$. Our theory can
be easily generalized to nonzero isospin chemical potential, $\mu_{\rm I}\neq0$. A large isospin chemical potential
leads to the Bose-Einstein condensation of charged pions and the BEC-BCS crossover \cite{Son2001,He2005}.

Introducing two auxiliary fields $\sigma$ and $\mbox{\boldmath{$\pi$}}$ which satisfy
equations of motion $\sigma=-2G\bar{\psi}\psi,\mbox{\boldmath{$\pi$}}=-2G\bar{\psi}
i\gamma_5\mbox{\boldmath{$\tau$}}\psi$, and applying the Hubbard-Strotonovich
transformation, we obtain
\begin{eqnarray}
{\cal Z}_{\rm NJL}=\int [d\psi] [d\bar{\psi}] [d\sigma]
[d\mbox{\boldmath{$\pi$}}]\exp\Big\{-{\cal
S}[\psi,\bar{\psi},\sigma,\mbox{\boldmath{$\pi$}}]\Big\},
\end{eqnarray}
where the action is given by
\begin{eqnarray}
{\cal S}[\psi,\bar{\psi},\sigma,\mbox{\boldmath{$\pi$}}]=\int dx
\frac{\sigma^2(x)+\mbox{\boldmath{$\pi$}}^2(x)}{4G}-\int dx\int
dx^\prime\bar{\psi}(x){\bf G}^{-1}(x,x^\prime)\psi(x^\prime),
\end{eqnarray}
with the inverse of the fermion Green's function
\begin{eqnarray}
{\bf
G}^{-1}(x,x^\prime)=\left\{\gamma^0(-\partial_\tau+\mu)+i\mbox{\boldmath{$\gamma$}}\cdot\mbox{\boldmath{$\nabla$}}-m_0
-\left[\sigma(x)+i\gamma_5\mbox{\boldmath{$\tau$}}\cdot\mbox{\boldmath{$\pi$}}(x)\right]\right\}\delta(x-x^\prime).
\end{eqnarray}
We integrate out the quark field and obtain
\begin{eqnarray}
{\cal Z}_{\rm NJL}&=&\int [d\sigma][d\mbox{\boldmath{$\pi$}}] \exp\left\{-{\cal S}_{\rm{eff}}[\sigma,\mbox{\boldmath{$\pi$}}]\right\},\nonumber\\
{\cal S}_{\rm{eff}}[\sigma,\mbox{\boldmath{$\pi$}}]&=&\int
dx\frac{\sigma^2(x)+\mbox{\boldmath{$\pi$}}^2(x)}{4G}-{\rm Tr}\ln{\bf
G}^{-1}(x,x^\prime).
\end{eqnarray}
At low temperature, we expect that the DCSB persists and we set $\langle\sigma(x)\rangle=\upsilon$ and 
$\langle\mbox{\boldmath{$\pi$}}(x)\rangle=0$. Applying again the field shifts
$\sigma(x)\rightarrow\upsilon+\sigma(x)$ and $\mbox{\boldmath{$\pi$}}(x)\rightarrow{\bf
0}+\mbox{\boldmath{$\pi$}}(x)$, we expand the effective action ${\cal
S}_{\rm{eff}}[\sigma,\mbox{\boldmath{$\pi$}}]$ in powers of the
fluctuations $\sigma(x)$ and $\mbox{\boldmath{$\pi$}}(x)$ and obtain
\begin{equation}
{\cal S}_{\rm{eff}}[\sigma,\mbox{\boldmath{$\pi$}}]={\cal S}_{\rm{eff}}^{(0)}+{\cal
S}_{\rm{eff}}^{(1)}[\sigma,\mbox{\boldmath{$\pi$}}] +{\cal
S}_{\rm{eff}}^{(2)}[\sigma,\mbox{\boldmath{$\pi$}}]+\cdots.
\end{equation}
In this work, we neglect the mesonic fluctuations higher than Gaussian.  The linear term
${\cal S}_{\rm{eff}}^{(1)}[\sigma,\mbox{\boldmath{$\pi$}}]$ can be shown to vanish.  
The partition function in this Gaussian approximation is given by
\begin{eqnarray}
{\cal Z}_{\rm NJL}\approx\exp\left\{-{\cal S}_{\rm{eff}}^{(0)}\right\}\int
[d\sigma][d\mbox{\boldmath{$\pi$}}] \exp\left\{-{\cal
S}_{\rm{eff}}^{(2)}[\sigma,\mbox{\boldmath{$\pi$}}]\right\}.
\end{eqnarray}
It is clear that the advantage of this Gaussian approximation is that we can complete the path integral
over the fluctuation fields $\sigma(x)$ and $\mbox{\boldmath{$\pi$}}(x)$. The
thermodynamic potential $\Omega=-\ln {\cal Z}_{\rm NJL}/(\beta V)$ is given by
\begin{eqnarray}
\Omega\approx\Omega_{\rm MF}+\Omega_{\rm FL},
\end{eqnarray}
where the mean-field contribution reads
\begin{eqnarray}
\Omega_{\rm MF}=\frac{1}{\beta V}{\cal S}_{\rm{eff}}^{(0)},
\end{eqnarray}
and the meson-fluctuation contribution  is given by
\begin{eqnarray}
\Omega_{\rm FL}=-\frac{1}{\beta V}\ln\left[\int [d\sigma][d\mbox{\boldmath{$\pi$}}]
\exp\left\{-{\cal S}_{\rm{eff}}^{(2)}[\sigma,\mbox{\boldmath{$\pi$}}]\right\}\right].
\end{eqnarray}

\subsection{Thermodynamics in mean-field approximation and phase diagram}\label{s3-1}

At finite temperature, the mean-field part ${\cal S}_{\rm{eff}}^{(0)}={\cal
S}_{\rm{eff}}[\upsilon,{\bf 0}]$ is given by
\begin{eqnarray}
{\cal S}_{\rm{eff}}^{(0)}=\beta V\frac{\upsilon^2}{4G}-\sum_{n}\sum_{\bf k}\ln\det
\left[{\cal G}^{-1}(ik_n,{\bf k})\right],
\end{eqnarray}
where
\begin{eqnarray}
{\cal G}^{-1}(ik_n,{\bf k})=(ik_n+\mu)\gamma^0-\mbox{\boldmath{$\gamma$}}\cdot{\bf k}-M
\end{eqnarray}
is the inverse of the mean-field quark Green's function in momentum space, with $k_n=(2n+1)\pi
T$  ($n\in\mathbb{Z}$) being the fermion Matsubara frequency and $M=m_0+\upsilon$ being the
effective quark mass. The mean-field thermodynamic potential can be evaluated as
\begin{eqnarray}
\Omega_{\rm MF}=\frac{\upsilon^2}{4G}-2N_cN_f\int\frac{d^3{\bf
k}}{(2\pi)^3}\left\{E_{\bf k}+\frac{1}{\beta}\ln\left[1+e^{-\beta (E_{\bf
k}-\mu)}\right]+\frac{1}{\beta}\ln\left[1+e^{-\beta (E_{\bf k}+\mu)}\right]\right\},
\end{eqnarray}
where $E_{\bf k}=\sqrt{{\bf k}^2+M^2}$.  As in the zero temperature case, we also regularize the integral over the quark momentum ${\bf k}$ via a hard cutoff $\Lambda$ ($|{\bf k}|<\Lambda$). The chiral condensate $\upsilon$
is determined by minimizing ${\cal S}_{\rm{eff}}^{(0)}$, i.e., $\partial {\cal
S}_{\rm{eff}}^{(0)}/\partial\upsilon=0$, leading to the  gap equation
\begin{eqnarray}\label{NJL-gapT}
M-m_0=4GN_cN_fM\int\frac{d^3{\bf k}}{(2\pi)^3}\frac{1-f(E_{\bf k}-\mu)-f(E_{\bf
k}+\mu)}{E_{\bf k}}.
\end{eqnarray}
Here $f(E)=1/(1+e^{\beta E})$ is the Fermi-Dirac distribution. If the phase transition is of first order,
the gap equation has multiple solutions. In this case, we compare their grand potentials and find the physical solution of $\upsilon$.

\begin{figure}
\centering
\includegraphics[width=3.5in]{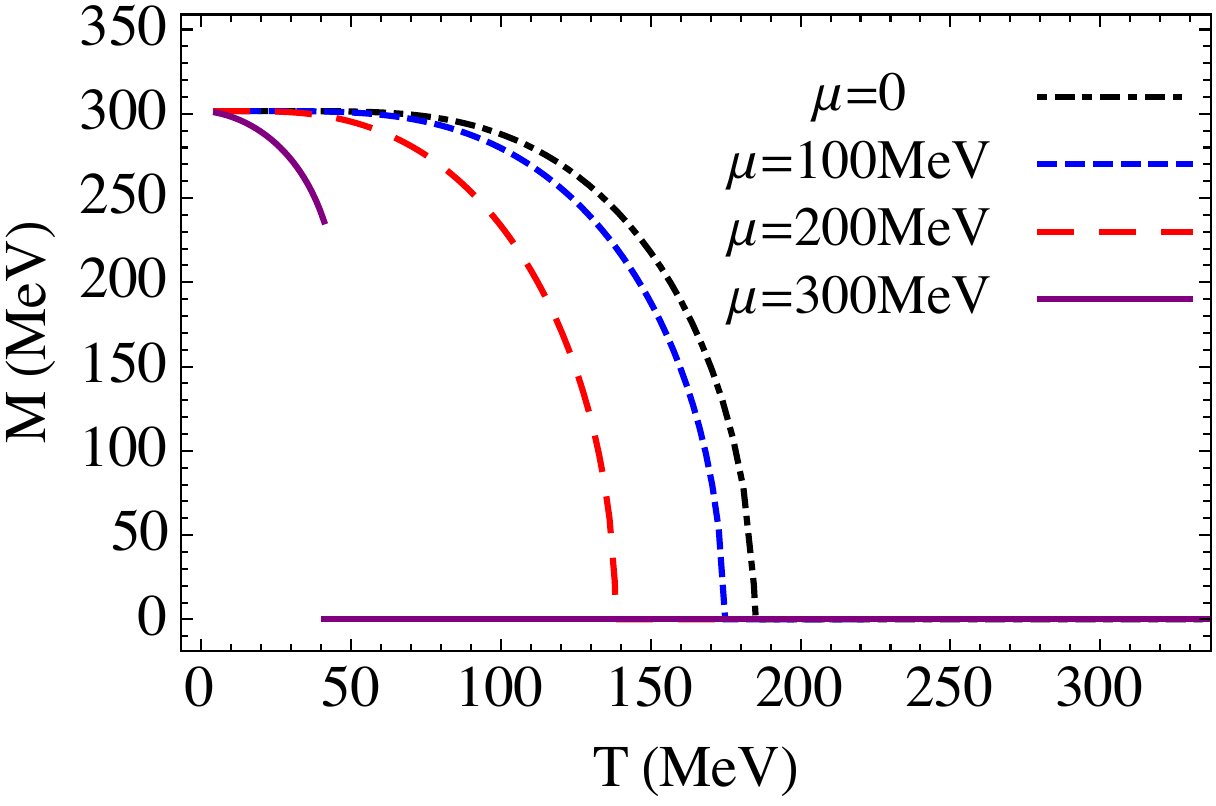}
\caption{ The effective quark mass $M$ as a function of $T$ for various values of the chemical potential $\mu$ in the chiral limit ($m_0=0$).} \label{fig1}
\end{figure}

\begin{figure}
\centering
\includegraphics[width=3.5in]{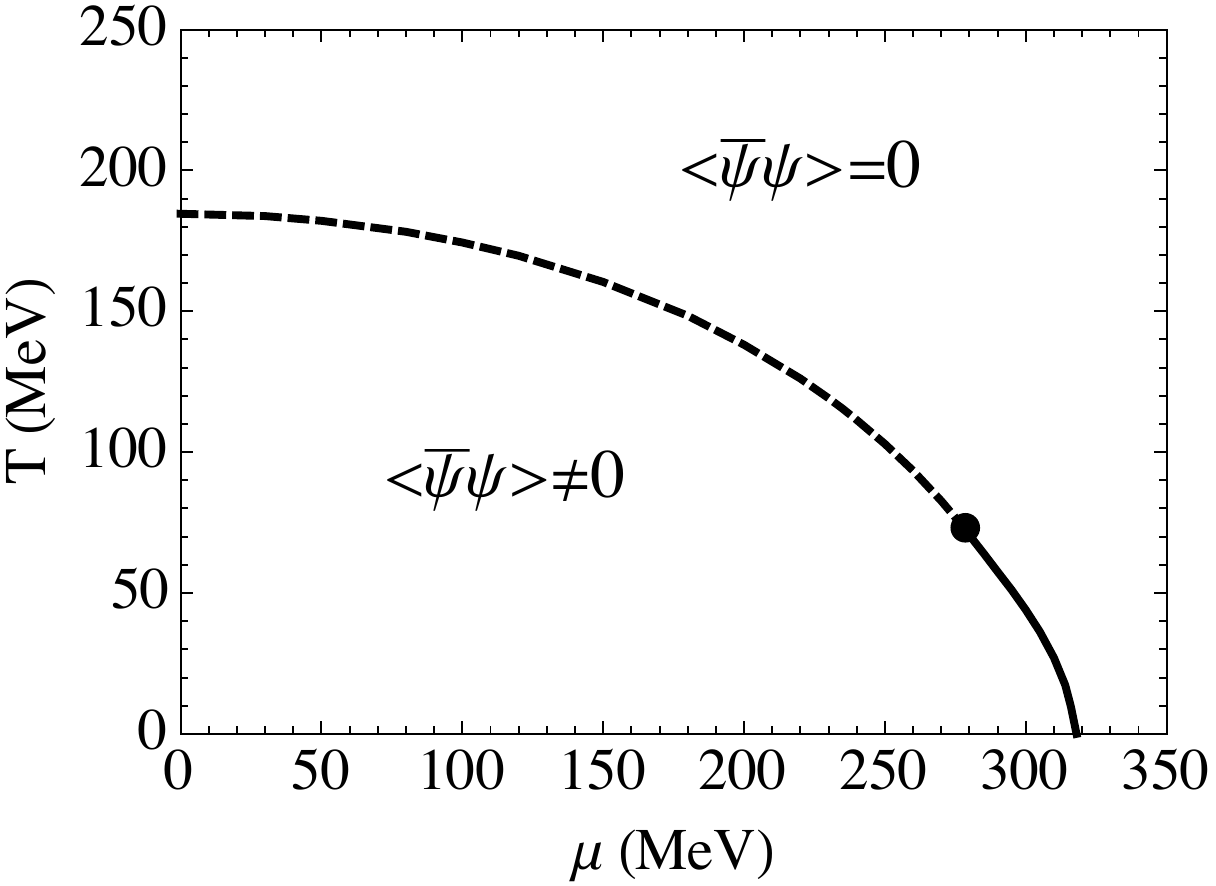}
\caption{The phase diagram of the NJL model in the $T$-$\mu$ plane for the chiral limit ($m_0=0$). The chiral symmetry broken and restored phases are denoted by 
$\langle\bar{\psi}\psi\rangle\neq0$ and  $\langle\bar{\psi}\psi\rangle=0$, respectively. The dashed and the solid lines represent the second-order and the first-order phase transitions, 
respectively.} \label{fig2}
\end{figure}

Figure \ref{fig1} shows the effective quark mass $M$ as a function of $T$ for various values of the chemical potential $\mu$ in the chiral limit ($m_0=0$). Figure \ref{fig2} shows the well-known phase diagram 
of the NJL model in the $T$-$\mu$ plane.  At small chemical potential, the chiral phase transition is of second order. 
It becomes of first order at large $\mu$.  Hence a tricritical point appears. For physical current quark mass, the second-order phase transition turns to be a crossover and the tricritical point becomes a critical endpoint.

\subsection{Thermodynamics including mesonic contributions}\label{s3-2}

Now we include the mesonic degrees of freedom. To this end, we consider the excitations corresponding to the fluctuation fields $\sigma(x)$ and $\mbox{\boldmath{$\pi$}}(x)$. It is convenient to work
in the momentum space by defining the Fourier transformation
\begin{eqnarray}
\phi_{\rm m}(x)=\sum_Q \phi_{\rm m}(Q)e^{-iq_l\tau+i{\bf q}\cdot{\bf r}},\ \ \ \ \ {\rm m}=0,1,2,3,
\end{eqnarray}
where $\phi_0=\sigma$ and $\phi_{\rm i}=\pi_{\rm i}$ (${\rm i}=1,2,3$). Here
$Q\equiv(iq_l, {\bf q})$ with $q_l=2l\pi T$ ($l\in\mathbb{Z}$)  being the boson Matsubara frequency. The notation $\sum_Q=\sum_l\int\frac{d^3{\bf q}}{(2\pi)^3}$ will be used throughout.  In the
momentum space, the inverse of the quark Green's function ${\bf G}^{-1}$ reads
\begin{eqnarray}
{\bf G}^{-1}(K,K^\prime)={\cal G}^{-1}(K)\delta_{K,K^\prime}-\Sigma_{\rm FL}(K,K^\prime),
\end{eqnarray}
where $K=(ik_n, {\bf k})$ and
\begin{eqnarray}
\Sigma_{\rm FL}(K,K^\prime)=\sum_{\rm m=0}^3\Gamma_{\rm m}\phi_{\rm m}(K-K^\prime).
\end{eqnarray}
Here we have defined $\Gamma_0=1$ and $\Gamma_{\rm i}=i\gamma_5\tau_{\rm i}$ (${\rm
i}=1,2,3$). Applying the derivative expansion, we obtain
\begin{eqnarray}
{\cal S}_{\rm{eff}}^{(2)}[\sigma,\mbox{\boldmath{$\pi$}}]=\frac{\beta V}{2}\sum_{{\rm
m,n}=0}^3\sum_Q\phi_{\rm m}(-Q)[{\bf D}^{-1}(Q)]_{\rm mn}\phi_{\rm n}(Q),
\end{eqnarray}
where
\begin{eqnarray}
[{\bf D}^{-1}(Q)]_{\rm mn}=\frac{\delta_{\rm mn}}{2G}+\Pi_{\rm mn}(Q)
\end{eqnarray}
is the inverse of the meson Green's function. The polarization function $\Pi_{\rm mn}(Q)$ is defined as
\begin{eqnarray}
\Pi_{\rm mn}(Q)=\frac{1}{\beta V}\sum_{K}{\rm Tr}\left[{\cal G}(K)\Gamma_{\rm m}{\cal
G}(K+Q)\Gamma_{\rm n}\right].
\end{eqnarray}
The notation $\sum_K=\sum_n\int\frac{d^3{\bf k}}{(2\pi)^3}$ will be used throughout. Since we consider the case $\mu_{\rm I}=0$, the off-diagonal components vanishes, i.e., 
$\Pi_{\rm mn}(Q)=\delta_{\rm mn}\Pi_{\rm m}(Q)$. It is also evident that $\Pi_{\rm m}(-Q)=\Pi_{\rm m}(Q)$.

The meson polarization functions $\Pi_{\rm m}(Q)$ can be evaluated as
\begin{eqnarray}
\Pi_{\rm 0}(iq_l,{\bf q})
&=&N_cN_f \int{d^3{\bf k}\over
(2\pi)^3}\Bigg[\left(\frac{1-f(E_{\bf k}^+)-f(E_{{\bf k}+{\bf q}}^-)}{iq_l-E_{\bf k}-E_{{\bf k}+{\bf q}}}-\frac{1-f(E_{\bf k}^-)-f(E_{{\bf k}+{\bf q}}^+)}{iq_l+E_{\bf k}+E_{{\bf k}+{\bf q}}}\right)
\left(1+\frac{{\bf k}\cdot ({\bf k+q})-M^2}{E_{\bf k} E_{{\bf k}+{\bf q}}}\right)\nonumber\\
&&+\left(\frac{f(E_{\bf k}^-)-f(E_{{\bf k}+{\bf q}}^-)}{iq_l+E_{\bf k}-E_{{\bf k}+{\bf q}}}-\frac{f(E_{\bf k}^+)-f(E_{{\bf k}+{\bf q}}^+)}{iq_l-E_{\bf k}+E_{{\bf k}+{\bf q}}}\right)
\left(1-\frac{{\bf k}\cdot ({\bf k+q})-M^2}{E_{\bf k} E_{{\bf k}+{\bf q}}}\right)\Bigg]
\end{eqnarray}
for ${\rm m}=0$, and 
\begin{eqnarray}
\Pi_{\rm m}(iq_l,{\bf q})
&=&N_cN_f \int{d^3{\bf k}\over
(2\pi)^3}\Bigg[\left(\frac{1-f(E_{\bf k}^+)-f(E_{{\bf k}+{\bf q}}^-)}{iq_l-E_{\bf k}-E_{{\bf k}+{\bf q}}}-\frac{1-f(E_{\bf k}^-)-f(E_{{\bf k}+{\bf q}}^+)}{iq_l+E_{\bf k}+E_{{\bf k}+{\bf q}}}\right)
\left(1+\frac{{\bf k}\cdot ({\bf k+q})+M^2}{E_{\bf k} E_{{\bf k}+{\bf q}}}\right)\nonumber\\
&&+\left(\frac{f(E_{\bf k}^-)-f(E_{{\bf k}+{\bf q}}^-)}{iq_l+E_{\bf k}-E_{{\bf k}+{\bf q}}}-\frac{f(E_{\bf k}^+)-f(E_{{\bf k}+{\bf q}}^+)}{iq_l-E_{\bf k}+E_{{\bf k}+{\bf q}}}\right)
\left(1-\frac{{\bf k}\cdot ({\bf k+q})+M^2}{E_{\bf k} E_{{\bf k}+{\bf q}}}\right)\Bigg]
\end{eqnarray}
for ${\rm m}=1,2,3$. Here we have defined $E_{\bf k}^\pm=E_{\bf k}\pm\mu$ for convenience. In the chiral limit, we can show that from the gap equation,  $1/(2G)+\Pi_{\rm m}(0,{\bf 0})=0$ (${\rm m}=1,2,3$) in the chiral symmetry broken phase $M\neq0$, which manifests the fact that the 
pions are Goldstone bosons in this phase. In the chiral symmetry restored phase,  $M=0$, we have obviously $\Pi_{\rm 0}(Q)=\Pi_{\rm 1}(Q)=\Pi_{\rm 2}(Q)=\Pi_{\rm 3}(Q)$, which indicates that the
sigma meson and the pions become degenerate.

In the Gaussian approximation, the path integral over $\phi_{\rm m}$ can be completed. The mesonic contribution to the thermodynamics potential can be evaluated as
\begin{eqnarray}
\Omega_{\rm FL}=\frac{1}{2\beta V}\sum_{Q}\ln\det\left[{\bf
D}^{-1}(Q)\right]=\frac{1}{2}\sum_{{\rm m}=0}^3\frac{1}{\beta}\sum_l\int\frac{d^3{\bf
q}}{(2\pi)^3}\ln\left[\frac{1}{2G}+\Pi_{\rm m}(iq_l,{\bf q})\right]e^{iq_l0^+}.
\end{eqnarray}
We can convert the summation over the boson Matsubara frequency to an contour
integration and obtain \cite{Hufner1994}
\begin{eqnarray}
\Omega_{\rm FL}=-\sum_{{\rm m}=0}^3\int\frac{d^3{\bf q}}{(2\pi)^3}\int_0^\infty\frac{d\omega}{2\pi i}
\left[\frac{\omega}{2}+\frac{1}{\beta}\ln\left(1-e^{-\beta\omega}\right)\right]
\frac{d}{d\omega}\ln\left[\frac{1+2G\Pi_{\rm m}(\omega+i0^+,{\bf q})}{1+2G\Pi_{\rm
m}(\omega-i0^+,{\bf q})}\right].
\end{eqnarray}
This result can be related to the Bethe-Uhlenbeck expression, i.e., the second virial contribution in terms of the two-body scattering phase shift \cite{Hufner1994}. We note that
$1+2G\Pi_{\rm m}(\omega+i0^+,{\bf q})$ is proportional to the $T$-matrix for the quark-antiquark scattering in the ${\rm m}$-channel, with total energy $\omega$
and momentum ${\bf q}$. The scattering matrix element can be written in the Jost representation as
 \begin{eqnarray}
{\cal S}_{\rm m}(\omega,{\bf q})=\frac{1+2G\Pi_{\rm m}(\omega-i0^+,{\bf q})}{1+2G\Pi_{\rm m}(\omega+i0^+,{\bf q})}.
\end{eqnarray}
The S-matrix element may has poles corresponding to mesonic bound states. Above the threshold for elastic scattering, it can be represented by a scattering phase shift as
 \begin{eqnarray}
{\cal S}_{\rm m}(\omega,{\bf q})=e^{2i\phi_{\rm m}(\omega,{\bf q})}.
\end{eqnarray}
Combining a possible pole term and the scattering contribution, we have \cite{Hufner1994}
\begin{eqnarray}
\Omega_{\rm FL}=\sum_{{\rm m}=0}^3\int\frac{d^3{\bf
q}}{(2\pi)^3}\int_0^\infty d\omega\left[\frac{\omega}{2}+\frac{1}{\beta}\ln\left(1-e^{-\beta\omega}\right)\right]
\left[\delta(\omega-\varepsilon_{\rm m}({\bf q}))+\frac{1}{\pi}\frac{\partial \phi_{\rm m}(\omega,{\bf q})}{\partial \omega}\right],
\end{eqnarray}
where the mesonic pole energy can be given by $\varepsilon_{\rm m}({\bf q})=\sqrt{{\bf q}^2+m_{\rm m}^2}$, with the in-medium meson mass $m_{\rm m}$. At low temperature, the above expression explicitly 
recovers the fact that the thermodynamic quantities are dominated by the lightest mesonic excitations, i.e., the pions \cite{Zhuang1994}. In the chiral limit, the pressure of the system at low temperature can be well given by
the pressure of a gas of noninteracting massless pions, $p=\pi^2T^4/30$.

\section{linear response theory in path integral}\label{s4}

We now start to study the linear response of the hot and dense matter in the NJL model, based on the description of the equilibrium thermodynamics in the last section. 
In this section, we introduce a generic theoretic framework to compute the following
imaginary-time-ordered current-current correlation function
\begin{equation}
\Pi^{\mu\nu}(\tau-\tau^\prime,{\bf r}-{\bf r}^\prime)=-\left\langle {\rm T}_\tau
\left[{\rm J}^\mu(\tau,{\bf r}){\rm J}^\nu(\tau^\prime,{\bf
r}^\prime)\right]\right\rangle_{\rm c},
\end{equation}
where ${\rm J}^\mu(\tau,{\bf r})$ can be any current operator. The notation $\langle
\cdots\rangle_{\rm c} $ denotes the connected piece of the correlation function. For a
pure fermionic field theory with a Lagrangian density
\begin{eqnarray}
{\cal L}[\psi,\bar{\psi}]=\bar\psi(i\gamma^\mu\partial_\mu-m_0)\psi+{\cal L}_{\rm
int}[\psi,\bar{\psi}],
\end{eqnarray}
the current operator is given by
\begin{eqnarray}
{\rm J}^{\mu}=\bar\psi\Gamma^\mu\psi,
\end{eqnarray}
where $\Gamma^\mu=\gamma^\mu\hat{X}$ with $\hat{X}$ being any Hermitian matrix in the
spin, flavor, and color spaces. For instance, the electromagnetic current is defined by $\hat{X}={\rm diag}(2e/3,-e/3)$ in the flavor space with $e$ being the elementary electric charge
and $\hat{X}=\gamma_5$ in the spin space gives the axial vector current.

Parallel to the path integral approach to the equilibrium thermodynamics, we introduce a path integral formalism for the linear response. In this formalism, we introduce an external source term to
compute the correlation function $\Pi^{\mu\nu}(\tau,{\bf r})$. The external source
physically represents an external perturbation applied to the system. The external
source here is actually an external gauge field $A_\mu(\tau,{\bf r})$ which couples to
the current ${\rm J}^\mu(\tau,{\bf r})$. We still use $x=(\tau, {\bf r})$ for
convenience. The partition function with external source is given by
\begin{eqnarray}
{\cal Z}[A] = \int [d\psi][d\bar{\psi}]\exp\left\{-{\cal S}[\psi,\bar{\psi};A]\right\},
\end{eqnarray}
where the action reads
\begin{eqnarray}
{\cal S}[\psi,\bar{\psi};A]=\int dx\ \bigg\{-{\cal L}[\psi,\bar{\psi}]-\mu\bar\psi\gamma^0\psi+A_\mu(x)\bar\psi\Gamma^\mu\psi\bigg\}.
\end{eqnarray}
It is convenient to use the generating functional ${\cal W}[A]$ defined as
\begin{eqnarray}
{\cal Z}[A] = \exp{\Big\{-{\cal W}[A]\Big\}}.
\end{eqnarray}
If the the generating functional can be computed exactly, the correlation function is given by
\begin{eqnarray}\label{Definition01}
\Pi^{\mu\nu}(\tau-\tau^\prime,{\bf r}-{\bf r}^\prime)=\frac{\delta^2{\cal W}[A]} {\delta
A_\mu(\tau,{\bf r})\delta A_\nu(\tau^\prime,{\bf r}^\prime)}\Bigg|_{A=0}.
\end{eqnarray}

In practice, we need to evaluate the generating functional in some approximations. It is convenient to work in the momentum space by making the Fourier transform
\begin{eqnarray}
A_\mu(x)=\sum_Q A_\mu(Q)e^{-iq_l\tau+i{\bf q}\cdot{\bf r}}.
\end{eqnarray}
To evaluate the correlation function, we expand the generating functional ${\cal W}[A]$ in powers of $A_\mu(Q)$. The expansion can be formally given by
\begin{eqnarray}
{\cal W}[A] = {\cal W}^{(0)}+{\cal W}^{(1)}[A]+{\cal W}^{(2)}[A]+\cdots,
\end{eqnarray}
where ${\cal W}^{(n)}$ is the $n$th-order expansion in $A_\mu(Q)$.  The zeroth-order 
contribution ${\cal W}^{(0)}$ recovers the equilibrium grand potential $\Omega$ with vanishing external source,
\begin{eqnarray}
{\cal W}^{(0)}=\beta V\Omega.
\end{eqnarray}
The first-order contribution ${\cal W}^{(1)}[A]$ gives nothing but the thermodynamic relation for
the charge density $n_{\rm X}=\langle {\rm J}^0\rangle$. We have $n_{\rm X}=-\partial\Omega/\partial\mu_{\rm X}$, where the chemical potential is defined as 
$\mu_{\rm X}=A_0(Q=0)$.  We have
\begin{eqnarray}
\frac{{\cal W}^{(1)}[A]}{\beta V}=-n_{\rm X}A_0(Q=0).
\end{eqnarray}
The second-order contribution ${\cal W}^{(2)}[A]$ characterizes the linear response. It can be formally given by
\begin{eqnarray}
\frac{{\cal W}^{(2)}[A]}{\beta V}=\frac{1}{2}\sum_Q \Pi^{\mu\nu}(Q)A_\mu(-Q)A_\nu(Q).
\end{eqnarray}
Here $\Pi^{\mu\nu}(Q)$ is just the correlation function in the momentum space. The static and long-wavelength limit of its $00$-component, $\Pi^{00}(Q=0)$, is related to
to the number susceptibility, i.e.,
\begin{eqnarray}
\lim_{{\bf q}\rightarrow0}\Pi^{00}(iq_l=0,{\bf q})=\frac{\partial^2\Omega(T,\mu_{\rm X})}{\partial \mu_{\rm X}^2}.
\end{eqnarray}
For instance, for the vector current with $\hat{X}=1$, $\Pi^{00}(Q=0)$ is proportional to the baryon number susceptibility. We note that the above
discussions are precise of the generating functional ${\cal W}[A]$ or its second-order expansion ${\cal W}^{(2)}[A]$ can be computed exactly. 

Now we turn to the NJL model. The partition function of the NJL model with external source is given by
\begin{eqnarray}
{\cal Z}_{\rm NJL}[A] = \int [d\psi][d\bar{\psi}]\exp\left\{-{\cal S}[\psi,\bar{\psi};A]\right\},
\end{eqnarray}
with the action
\begin{eqnarray}
{\cal S}[\psi,\bar{\psi};A]=\int dx\ \bigg\{-{\cal L}_{\rm
NJL}[\psi,\bar{\psi}]-\mu\bar\psi\gamma^0\psi+A_\mu(x)\bar\psi\Gamma^\mu\psi\bigg\}.
\end{eqnarray}
Again, introducing two auxiliary fields $\sigma$ and $\mbox{\boldmath{$\pi$}}$ which satisfy
equations of motion
$\sigma=-2G\bar{\psi}\psi,\mbox{\boldmath{$\pi$}}=-2G\bar{\psi}i\gamma_5\mbox{\boldmath{$\tau$}}\psi$,
and applying the Hubbard-Strotonovich transformation, we obtain
\begin{eqnarray}
{\cal Z}_{\rm NJL}[A]=\int [d\psi] [d\bar{\psi}] [d\sigma]
[d\mbox{\boldmath{$\pi$}}]\exp\left\{-{\cal
S}[\psi,\bar{\psi},\sigma,\mbox{\boldmath{$\pi$}};A]\right\},
\end{eqnarray}
where the action now reads
\begin{eqnarray}
{\cal S}[\psi,\bar{\psi},\sigma,\mbox{\boldmath{$\pi$}};A]=\int dx
\frac{\sigma^2(x)+\mbox{\boldmath{$\pi$}}^2(x)}{4G}-\int dx\int
dx^\prime\bar{\psi}(x){\bf G}_{\rm A}^{-1}(x,x^\prime)\psi(x^\prime),
\end{eqnarray}
with the inverse of the fermion Green's function
\begin{eqnarray}
{\bf G}_{\rm
A}^{-1}(x,x^\prime)=\Big\{\gamma^0(-\partial_\tau+\mu)+i\mbox{\boldmath{$\gamma$}}\cdot\mbox{\boldmath{$\nabla$}}-m_0
-\left[\sigma(x)+i\gamma_5\mbox{\boldmath{$\tau$}}\cdot\mbox{\boldmath{$\pi$}}(x)\right]-\Gamma^\mu
A_\mu(x)\Big\}\delta(x-x^\prime).
\end{eqnarray}
Integrating out the quark field leads to
\begin{eqnarray}
{\cal Z}_{\rm NJL}[A]&=&\int [d\sigma][d\mbox{\boldmath{$\pi$}}] \exp\left\{-{\cal S}_{\rm{eff}}[\sigma,\mbox{\boldmath{$\pi$}};A]\right\},\nonumber\\
{\cal S}_{\rm{eff}}[\sigma,\mbox{\boldmath{$\pi$}};A]&=&\int
dx\frac{\sigma^2(x)+\mbox{\boldmath{$\pi$}}^2(x)}{4G}-{\rm Tr}\ln{\bf G}_{\rm
A}^{-1}(x,x^\prime).
\end{eqnarray}

The treatment of the the expectation values of the meson fields $\sigma(x)$ and
$\mbox{\boldmath{$\pi$}}(x)$, or their \emph{classical fields} $\sigma_{\rm cl}(x)$ and
$\mbox{\boldmath{$\pi$}}_{\rm cl}(x)$, becomes nontrivial. In the absence of the external
source, we have chosen $\sigma_{\rm cl}(x)=\upsilon$ and $\mbox{\boldmath{$\pi$}}_{\rm
cl}(x)={\bf 0}$, which are static and homogeneous.  However, in the presence of the external
source, they are generally no longer static and homogeneous.  Again, we apply the
field shifts, $\sigma(x)\rightarrow\sigma_{\rm cl}(x)+\sigma(x)$ and
$\mbox{\boldmath{$\pi$}}(x)\rightarrow\mbox{\boldmath{$\pi$}}_{\rm
cl}(x)+\mbox{\boldmath{$\pi$}}(x)$, and expand the effective action ${\cal
S}_{\rm{eff}}[\sigma,\mbox{\boldmath{$\pi$}}]$ in powers of the
fluctuations $\sigma(x)$ and $\mbox{\boldmath{$\pi$}}(x)$. We obtain
\begin{equation}
{\cal S}_{\rm{eff}}[\sigma,\mbox{\boldmath{$\pi$}};A]={\cal S}_{\rm{eff}}^{(0)}[A]+{\cal
S}_{\rm{eff}}^{(1)}[\sigma,\mbox{\boldmath{$\pi$}};A] +{\cal
S}_{\rm{eff}}^{(2)}[\sigma,\mbox{\boldmath{$\pi$}};A]+\cdots.
\end{equation}
Parallel to the case without external source, we neglect the mesonic fluctuations higher
than Gaussian. The linear term  ${\cal S}_{\rm{eff}}^{(1)}[\sigma,\mbox{\boldmath{$\pi$}};A]$ can be shown to vanish once the
classical fields $\sigma_{\rm cl}(x)$ and $\mbox{\boldmath{$\pi$}}_{\rm cl}(x)$ are
determined by minimizing ${\cal S}_{\rm{eff}}^{(0)}[A]$. 
The partition function in the Gaussian approximation is given by
\begin{eqnarray}
{\cal Z}_{\rm NJL}\approx\exp\left\{-{\cal S}_{\rm{eff}}^{(0)}[A]\right\}\int
[d\sigma][d\mbox{\boldmath{$\pi$}}] \exp\left\{-{\cal
S}_{\rm{eff}}^{(2)}[\sigma,\mbox{\boldmath{$\pi$}};A]\right\}.
\end{eqnarray}
Therefore, in this Gaussian approximation, the generating functional
${\cal W}_{\rm NJL}[A]$ includes both the mean-field (MF) and the meson-fluctuation
(FL) contributions. We have
\begin{eqnarray}
{\cal W}_{\rm NJL}[A] = {\cal W}_{\rm MF}[A]+{\cal W}_{\rm FL}[A],
\end{eqnarray}
where
\begin{eqnarray}
{\cal W}_{\rm MF}[A]={\cal S}_{\rm{eff}}^{(0)}[A],\ \ \ \ {\cal W}_{\rm FL}[A]=-\ln\left[\int [d\sigma][d\mbox{\boldmath{$\pi$}}]
\exp\left\{-{\cal S}_{\rm{eff}}^{(2)}[\sigma,\mbox{\boldmath{$\pi$}};A]\right\}\right].
\end{eqnarray}
It is clear that in the path integral, we can treat the equilibrium thermodynamics and the linear response at the same footing. The mean-field and the meson-fluctuation
contributions to the generating functional can be expanded in powers of the external source as
\begin{eqnarray}
{\cal W}_{\rm MF}[A] &=& {\cal W}_{\rm MF}^{(0)}+{\cal W}_{\rm MF}^{(1)}[A]+{\cal W}_{\rm MF}^{(2)}[A]+\cdots,\nonumber\\
{\cal W}_{\rm FL}[A] &=& {\cal W}_{\rm FL}^{(0)}+{\cal W}_{\rm FL}^{(1)}[A]+{\cal W}_{\rm FL}^{(2)}[A]+\cdots.
\end{eqnarray}
It is evident that the zeroth-order contributions recover the equilibrium thermodynamic potentials, i.e.,  ${\cal W}_{\rm MF}^{(0)}=\beta V \Omega_{\rm MF}$ and 
${\cal W}_{\rm FL}^{(0)}=\beta V \Omega_{\rm FL}$.

So far the dependence on the classical fields $\sigma_{\rm cl}(x)$ and
$\mbox{\boldmath{$\pi$}}_{\rm cl}(x)$ is not explicitly shown. They are not independent
quantities and should be determined as functional of the external source via some gap equations. Now we write
\begin{eqnarray}
{\cal W}_{\rm NJL}[A] = {\cal W}_{\rm MF}[A;\sigma_{\rm cl},
\mbox{\boldmath{$\pi$}}_{\rm cl}]+{\cal W}_{\rm FL}[A;\sigma_{\rm cl},
\mbox{\boldmath{$\pi$}}_{\rm cl}].
\end{eqnarray}
Parallel to the theory of the equilibrium thermodynamics, we require that the classical fields are
determined by minimizing the \emph{mean-field} part of the generating functional, i.e.,
\begin{equation}\label{LRT-GAP}
\frac{\delta{\cal W}_{\rm MF}[A;\sigma_{\rm cl}, \mbox{\boldmath{$\pi$}}_{\rm
cl}]}{\delta\sigma_{\rm cl}(x)}=0,\ \ \ \ \ \ \frac{\delta{\cal W}_{\rm
MF}[A;\sigma_{\rm cl}, \mbox{\boldmath{$\pi$}}_{\rm
cl}]}{\delta\mbox{\boldmath{$\pi$}}_{\rm cl}(x)}=0.
\end{equation}
Once this extreme condition is imposed, we can show that the linear term  ${\cal
S}_{\rm{eff}}^{(1)}[\sigma,\mbox{\boldmath{$\pi$}};A]$ vanishes exactly. Moreover, it is also necessary to maintain the
Goldstone's theorem. Solving the extreme condition formally, we have
\begin{eqnarray}
\sigma_{\rm cl}(x)=F_\sigma[A],\ \ \ \ \ \ \mbox{\boldmath{$\pi$}}_{\rm cl}(x)={\bf F}_\pi[A].
\end{eqnarray}
Substituting these solutions into the generating functional, we finally eliminate the
dependence on the classical fields.

\section{Linear response in mean-field Theory: Random Phase approximation}\label{s5}
We first the linear response in the mean-field approximation, i.e.,  ${\cal W}_{\rm NJL}[A] \simeq {\cal
W}_{\rm MF}[A]$. We will see that the response functions in this approximation recovers the famous random
phase approximation (RPA) developed in early condensed matter theory. Since we are interested in the response to an infinitesimal
external source, we expect that the induced perturbations to the classical fields are also
infinitesimal. Therefore, we have
\begin{equation}
\sigma_{\rm cl}(x)=\upsilon+\eta_0(x),\ \ \ \ \ \mbox{\boldmath{$\pi$}}_{\rm cl}(x)={\bf
0}+\mbox{\boldmath{$\eta$}}(x),
\end{equation}
where the static and uniform part $\upsilon$ is the chiral condensate with vanishing external source. The generating functional in the mean-field approximation is given by
\begin{eqnarray}
{\cal W}_{\rm MF}[A;\sigma_{\rm cl}, \mbox{\boldmath{$\pi$}}_{\rm cl}]= \int dx
\frac{\sigma_{\rm cl}^2(x)+\mbox{\boldmath{$\pi$}}^2_{\rm cl}(x)}{4G} - \mbox{Trln}\left[{\cal G}_{\rm A}^{-1}(x,x^\prime)\right].
\end{eqnarray}
Here ${\cal G}_{\rm A}^{-1}$ is the inverse of the fermion Green's function in the mean-field approximation with external source. It can be expressed as
\begin{equation}
{\cal G}_{\rm A}^{-1}(x,x^\prime)={\cal G}^{-1}(x,x^\prime)-\Sigma_{\rm A}(x,x^\prime),
\end{equation}
where the two terms are defined as
\begin{eqnarray}
&&{\cal G}^{-1}(x,x^\prime)=\left[\gamma^0(-\partial_\tau+\mu)+i\mbox{\boldmath{$\gamma$}}\cdot\mbox{\boldmath{$\nabla$}}-M\right]\delta(x-x^\prime),\nonumber\\
&&\Sigma_{\rm A}(x,x^\prime)=\left[\sum_{\rm m=0}^3\Gamma_{\rm m}\eta_{\rm
m}(x)+\Gamma^\mu A_\mu(x)\right]\delta(x-x^\prime).
\end{eqnarray}
Here $M=m_0+\upsilon$ is the effective quark mass as we have defined in the absence of the external source.

Now we turn to the momentum space via the Fourier transform
\begin{eqnarray}
\eta_{\rm m}(x)=\sum_Q \eta_{\rm m}(Q)e^{-iq_l\tau+i{\bf q}\cdot{\bf r}}.
\end{eqnarray}
In the momentum space, the inverse of the fermion Green's function is given by
\begin{equation}
{\cal G}_{\rm A}^{-1}(K,K^\prime)={\cal G}^{-1}(K)\delta_{K,K^\prime}-\Sigma_{\rm A}(K,K^\prime),
\end{equation}
where ${\cal G}^{-1}(K)$ is given by (33) and
\begin{eqnarray}
\Sigma_{\rm A}(K,K^\prime)=\sum_{\rm m=0}^3\Gamma_{\rm m}\eta_{\rm
m}(K-K^\prime)+\Gamma^\mu A_\mu(K-K^\prime).
\end{eqnarray}
By using the derivative expansion, we can expand the generating functional in powers of the external source as well as the induced
perturbations $\eta_{\rm m}$. We have 
\begin{eqnarray}
{\cal W}_{\rm MF}[A;\eta] = {\cal W}_{\rm MF}^{(0)}+{\cal W}_{\rm MF}^{(1)}[A;\eta]+{\cal W}_{\rm MF}^{(2)}[A;\eta]+\cdots,
\end{eqnarray}
where it is obvious that ${\cal W}_{\rm MF}^{(0)}=\beta V\Omega_{\rm MF}$. Note that the induced perturbation should be finally eliminated  via the
gap equation (\ref{LRT-GAP}).

The linear term ${\cal W}_{\rm MF}^{(1)}[A;\eta]$ can be evaluated as
\begin{eqnarray}
\frac{{\cal W}_{\rm MF}^{(1)}[A;\eta]}{\beta V}=\left[\frac{\upsilon}{2G}+\frac{1}{\beta
V}\sum_K{\rm Tr}{\cal G}(K)\right]\eta_0(0) +\frac{1}{\beta V}\sum_K{\rm Tr}\left[{\cal
G}(K)i\gamma^5\mbox{\boldmath{$\tau$}}\right]\cdot\mbox{\boldmath{$\eta$}}(0)
+\frac{1}{\beta V}\sum_K{\rm Tr}\left[{\cal G}(K)\Gamma^\mu\right] A_\mu(0).
\end{eqnarray}
It is related only to the $Q=0$ component of the external source and the induced perturbations. The explicit form of ${\cal G}(K)$ can be
evaluated as
\begin{eqnarray}
{\cal G}(K)=\frac{1}{ik_n-E_{\bf k}} \Lambda_{+}({\bf k})\gamma_0 + \frac{1}{ik_n+E_{\bf k}}
\Lambda_{-}({\bf k})\gamma_0,
\end{eqnarray}
where the the energy projectors  $\Lambda_\pm({\bf k})$ are given by
\begin{equation}
\Lambda_{\pm}({\bf k}) = \frac{1}{2}\left[1\pm{\gamma_0\left(\mbox{\boldmath{$\gamma$}}\cdot{\bf k}+M\right)\over E_{\bf k}} \right].
\end{equation}
Using the gap equation (35), we can show that the only nonvanishing part is
related to the number density, i.e.,
\begin{eqnarray}
\frac{{\cal W}_{\rm MF}^{(1)}[A;\eta]}{\beta V}=-(n_{\rm X})_{\rm MF}A_0(0),
\end{eqnarray}
where the number density is given by
\begin{eqnarray}
(n_{\rm X})_{\rm MF}=\frac{1}{\beta V}\sum_K{\rm Tr}\left[{\cal G}(K)\Gamma^0\right].
\end{eqnarray}
It is evident that $(n_{\rm X})_{\rm MF}=-\partial\Omega_{\rm MF}/\partial\mu_{\rm X}$ with the chemical potential $\mu_{\rm X}=A_0(0)$.

The linear response is characterized by the quadratic term ${\cal W}_{\rm
MF}^{(2)}[A;\eta]$. By making use of the derivative expansion and completing the trace in the
momentum space, we obtain
\begin{eqnarray}
{\cal W}_{\rm MF}^{(2)}[A;\eta]=\frac{\beta V}{4G}\sum_{\rm
m=0}^3\sum_Q\eta_{\rm m}(-Q)\eta_{\rm m}(Q)+\frac{1}{2}\sum_{K}\sum_{K^\prime}{\rm Tr}
\left[{\cal G}(K)\Sigma_{\rm A}(K,K^\prime){\cal G}(K^\prime)\Sigma_{\rm
A}(K^\prime,K)\right]
\end{eqnarray}
Defining $Q=K^\prime-K$, we obtain
\begin{eqnarray}
\frac{{\cal W}_{\rm MF}^{(2)}[A;\eta]}{\beta V}&=&\frac{1}{2}\sum_{Q}\Pi_{\rm
b}^{\mu\nu}(Q)A_\mu(-Q)A_\nu(Q)
+\frac{1}{2}\sum_{\rm m=0}^3\sum_{Q}\left[\frac{1}{2G}+\Pi_{\rm m}(Q)\right]\eta_{\rm m}(-Q)\eta_{\rm m}(Q)\nonumber\\
&+&\sum_{\rm m=0}^3\sum_{Q}C^{\mu}_{\rm m}(Q)A_\mu(-Q)\eta_{\rm m}(Q).
\end{eqnarray}
Here the the bare response function $\Pi_{\rm b}^{\mu\nu}(Q)$  is defined as
\begin{eqnarray}
\Pi_{\rm b}^{\mu\nu}(Q)=\frac{1}{\beta V}\sum_{K}{\rm Tr}\left[{\cal
G}(K)\Gamma^\mu{\cal G}(K+Q)\Gamma^\nu\right]
\end{eqnarray}
and the coupling function $C^\mu_{\rm m}(Q)$ is given by
\begin{eqnarray}
C^\mu_{\rm m}(Q)&=&\frac{1}{\beta V}\sum_{K}{\rm Tr}\left[{\cal G}(K)\Gamma^\mu{\cal
G}(K+Q)\Gamma_{\rm m}\right].
\end{eqnarray}
The meson polarization functions $\Pi_{\rm m}(Q)$ have been given in Sec. \ref{s3}.

The final task is to eliminate the induced perturbations. For the purpose of linear response, the induced perturbations $\eta_{\rm m}(Q)$ can be determined by
\begin{equation}\label{RPA-GAP}
\frac{\delta{\cal W}_{\rm MF}^{(2)}[A;\eta]}{\delta \eta_{\rm m}(Q)}=0.
\end{equation}
Using the explicit form of ${\cal W}_{\rm MF}^{(2)}[A;\eta]$, we obtain
\begin{equation}\label{RPA-GAP}
\eta_{\rm m}(Q)=-\frac{C^\mu_{\rm m}(-Q)A_\mu(Q)}{\frac{1}{2G}+\Pi_{\rm m}(Q)}+O(A^2),\
\ \ \ \ \ \eta_{\rm m}(-Q)=-\frac{C^\mu_{\rm m}(Q)A_\mu(-Q)}{\frac{1}{2G}+\Pi_{\rm
m}(Q)}+O(A^2),
\end{equation}
where we have used the fact $\Pi_{\rm m}(-Q)=\Pi_{\rm m}(Q)$. Using the above results to
eliminate the induced perturbations, we finally obtain
\begin{eqnarray}
\frac{{\cal W}_{\rm MF}^{(2)}[A]}{\beta V}=\frac{1}{2}\sum_{Q}\Pi_{\rm
MF}^{\mu\nu}(Q)A_\mu(-Q)A_\nu(Q),
\end{eqnarray}
where the full response function in the mean-field theory reads
\begin{eqnarray}
\Pi_{\rm MF}^{\mu\nu}(Q)=\Pi_{\rm b}^{\mu\nu}(Q)-\sum_{\rm m=0}^3\frac{C^\mu_{\rm
m}(Q)C^\nu_{\rm m}(-Q)}{\frac{1}{2G}+\Pi_{\rm m}(Q)}.
\end{eqnarray}
This result recovers nothing but the quasi-particle random phase approximation widely used in condensed
matter theory \cite{He2016}. We note that in addition to the pure quasi-particle contribution
$\Pi_{\rm b}^{\mu\nu}(Q)$, the linear response can couple to the collective mesonic
modes once $C^\mu_{\rm m}(Q)\neq0$. In other words, the response function reveals the
meson properties and also possibly the phase transitions.

In the chiral limit ($m_0=0$), we can show that $C^\mu_{\rm m}(Q)=0$ in the chiral symmetry
restored phase ($T>T_c$). In this case, the quasi-particle random phase approximation
just describes the linear response of a hot and dense gas of noninteracting quarks. This is
obviously inadequate. We will discuss the linear response theory beyond the quasi-particle random
phase approximation in Sec. VII.

\section{Dynamical density responses in random phase approximation}\label{s6}

As an application of the mean-field theory or the random phase approximation, we study the linear responses to some density perturbations. To be specific, we consider the following
$\hat{X}$ operators: (1) $\hat{X}=1$, corresponding to the vector current; (2) $\hat{X}=\tau_3$, corresponding to the isospin vector current;
(3) $\hat{X}=\gamma_5$, corresponding to the axial vector current; (4) $\hat{X}=\tau_3\gamma_5$, corresponding to the isospin axial vector current.
The $0$-component of the current ${\rm J}^\mu$ is related to the baryon density, the isospin density, the axial baryon density, and the axial isospin density,
respectively. The density response function $\chi(iq_l,{\bf q})$ is given by the $00$-component of the
response function $\Pi^{\mu\nu}(Q)$. In the mean-field theory, it is given by
\begin{eqnarray}
\chi(iq_l,{\bf q})=\Pi_{\rm MF}^{00}(Q)=\Pi_{\rm b}^{00}(Q)-\sum_{\rm m=0}^3\frac{C^0_{\rm
m}(Q)C^0_{\rm m}(-Q)}{\frac{1}{2G}+\Pi_{\rm m}(Q)}.
\end{eqnarray}
In practice, we define the dynamic structure factor $S(\omega,{\bf q})$, which is
related to the density response function $\chi(iq_l,{\bf q})$ via the fluctuation-dissipation theorem. It is defined as
\begin{eqnarray}
S(\omega,{\bf q})=-\frac{1}{\pi}\frac{1}{1-e^{-\beta\omega}}{\rm
Im}\chi(\omega+i\epsilon,{\bf q}).
\end{eqnarray}
In the following we will be interested in the long-wavelength limit ${\bf q}=0$ and focus on the pure dynamical effect.

\subsection{Vector current}

For the vector current $\hat{X}=1$, the bare response function is given by
\begin{eqnarray}
\Pi_{\rm b}^{00}(Q)=\frac{1}{\beta V}\sum_{K}{\rm Tr}\left[{\cal G}(K)\gamma^0{\cal
G}(K+Q)\gamma^0 \right] 
\end{eqnarray}
At ${\bf q}=0$, we can show that $\Pi_{\rm b}^{00}(iq_l, {\bf q}=0)$ vanishes. The coupling function is given by
\begin{eqnarray}
C^0_{\rm m}(Q)=\frac{1}{\beta V}\sum_{K}{\rm Tr}\left[{\cal G}(K)\gamma^0{\cal G}(K+Q)\Gamma_{\rm m}\right].
\end{eqnarray}
At ${\bf q}=0$, we can show that $C^0_{\rm m}(iq_l, {\bf q}=0)$ vanish for all ${\rm m}=0,1,2,3$. Therefore, for the baryon density response, the dynamic structure factor vanishes
at ${\bf q}=0$, i.e.,
\begin{eqnarray}
S(\omega,{\bf q}=0)=0
\end{eqnarray}

\subsection{Isospin vector current}

For the isospin vector current $\hat{X}=\tau_3$, the bare response function is given by
\begin{eqnarray}
\Pi_{\rm b}^{00}(Q)=\frac{1}{\beta V}\sum_{K}{\rm Tr}\left[{\cal G}(K)\gamma^0\tau_3{\cal G}(K+Q)\gamma^0 \tau_3\right]. 
\end{eqnarray}
At ${\bf q}=0$, we can show that $\Pi_{\rm b}^{00}(iq_l, {\bf q}=0)$ vanishes. The coupling function is given by
\begin{eqnarray}
C^0_{\rm m}(Q)=\frac{1}{\beta V}\sum_{K}{\rm Tr}\left[{\cal G}(K)\gamma^0\tau_3{\cal G}(K+Q)\Gamma_{\rm m}\right].
\end{eqnarray}
At ${\bf q}=0$, we can show that $C^0_{\rm m}(iq_l, {\bf q}=0)$ vanish for all ${\rm m}=0,1,2,3$.  Therefore, for the isospin density response, the dynamic structure factor also vanishes
at ${\bf q}=0$, i.e.,
\begin{eqnarray}
S(\omega,{\bf q}=0)=0
\end{eqnarray}

\subsection{Axial vector current}

\begin{figure}
\centering
\includegraphics[width=3.5in]{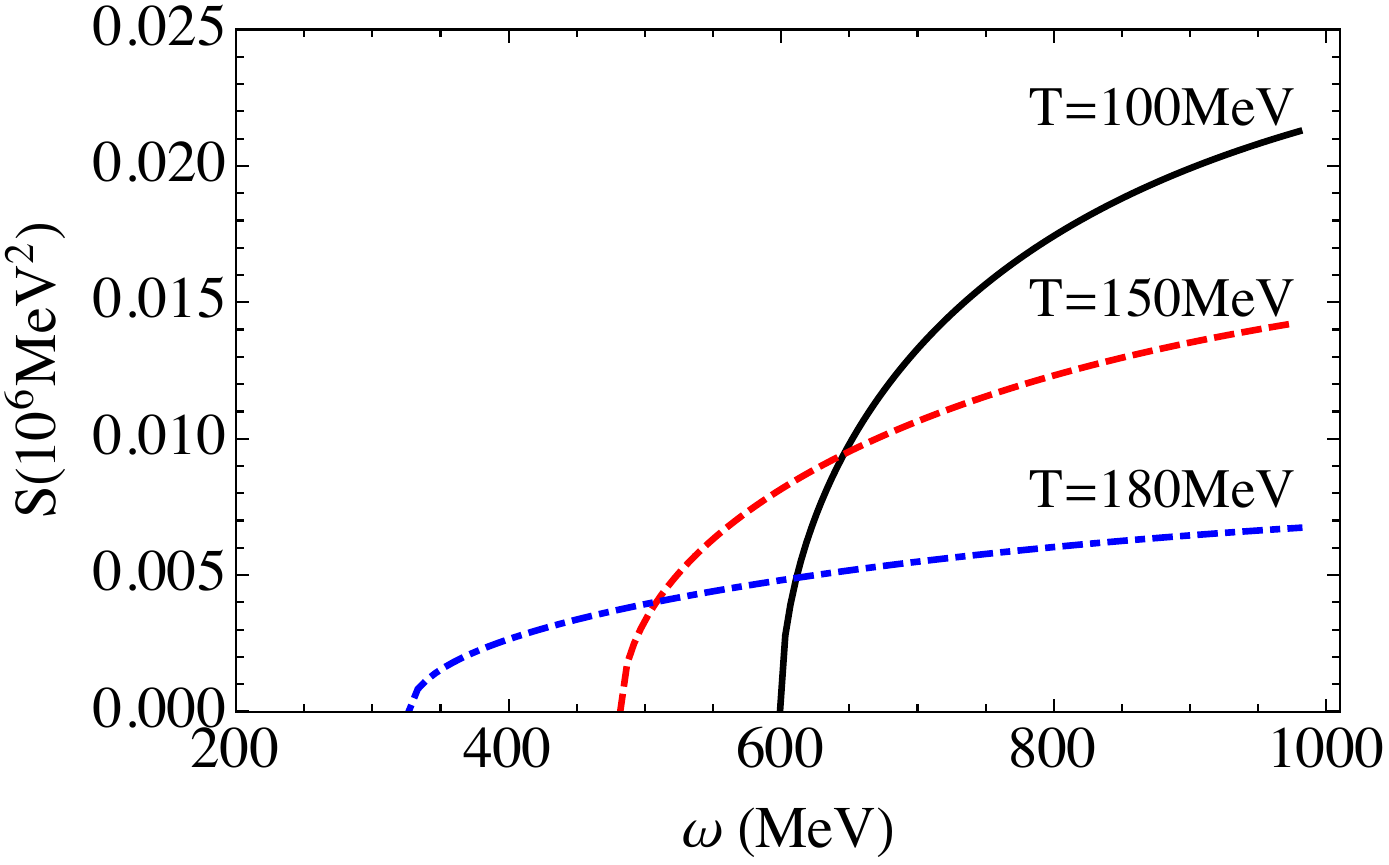}
\caption{ The dynamical structure factor $S(\omega,{\bf q})$ at ${\bf q}=0$ for the axial baryon density response ($\hat{X}=\gamma_5$) at various
values of $T$ and at $\mu=0$. We consider the physical current quark mass $m_0=5$MeV.} \label{fig3}
\end{figure}

For the axial vector current $\hat{X}=\gamma_5$, the bare response function is given by
\begin{eqnarray}
\Pi_{\rm b}^{00}(Q)=\frac{1}{\beta V}\sum_{K}{\rm Tr}\left[{\cal G}(K)\gamma^0\gamma_5{\cal G}(K+Q)\gamma^0 \gamma_5\right].
\end{eqnarray}
Completing the trace and the Matsubara sum, we obtain
\begin{eqnarray}
\Pi_{\rm b}^{00}(iq_l, {\bf q}=0)=2N_c N_f\int{d^3{\bf k}\over
(2\pi)^3} \frac{M^2}{E_{\bf k}^2}\left(\frac{1}{iq_l-2E_{\bf k}}-\frac{1}{iq_l+2E_{\bf k}}\right)\left[1-f(E_{\bf k}-\mu)-f(E_{\bf k}+\mu)\right].
\end{eqnarray}
The coupling function is given by
\begin{eqnarray}
C^0_{\rm m}(Q)=\frac{1}{\beta V}\sum_{K}{\rm Tr}\left[{\cal G}(K)\gamma^0
\gamma^5{\cal G}(K+Q)\Gamma_{\rm m}\right].
\end{eqnarray}
At ${\bf q}=0$, we can show that $C^0_{\rm m}(iq_l, {\bf q}=0)$ vanish for all ${\rm m}=0,1,2,3$. Therefore, the axial baryon density response has nonzero dynamical
structure factor at ${\bf q}=0$. It does not couple to the mesonic modes and is given by $\chi(iq_l,{\bf q}=0)=\Pi_{\rm b}^{00}(iq_l,{\bf q}=0)$. The dynamical structure factor reads
\begin{eqnarray}
S(\omega,{\bf q}=0)
 =N_cN_f \frac{M^2}{2\pi^2} \frac{ \sqrt{\omega^2-4M^2}}{\omega}\frac{\Theta(|\omega|-2M)}{1-e^{-\beta\omega}}
\left[1-\frac{1}{e^{\beta(\frac{1}{2}\omega-\mu)}+1}-\frac{1}{e^{\beta(\frac{1}{2}\omega+\mu)}+1}\right].
\end{eqnarray}
We note that similar result was also obtained in \cite{Hou2018}. It is evident that the dynamical structure factor for the axial baryon density response is a direct reflection of the quark mass gap.
$S(\omega,{\bf q}=0)$ is nonzero only for $|\omega|$ larger than twice of the quark mass gap. Figure \ref{fig3} shows the dynamical structure factor $S(\omega,{\bf q}=0)$ for various values of the temperature.
With increasing temperature, the threshold $\omega_{\rm th}=2M$ becomes smaller, and finally $\omega_{\rm th}\rightarrow 0$ in the high $T$ limit.

\subsection{Isospin axial vector current}

For the isospin axial vector current $\hat{X}=\tau_3\gamma^5$, the bare response function is given by
\begin{eqnarray}
\Pi_{\rm b}^{00}(Q)=\frac{1}{\beta V}\sum_{K}{\rm Tr}\left[{\cal G}(K)\gamma^0\tau_3\gamma^5{\cal G}(K+Q)\gamma^0 \tau_3\gamma^5\right].
\end{eqnarray}
Completing the trace and the Matsubara sum, we obtain
\begin{eqnarray}
\Pi_{\rm b}^{00}(iq_l, {\bf q}=0)=2N_c N_f\int{d^3{\bf k}\over
(2\pi)^3} \frac{M^2}{E_{\bf k}^2}\left(\frac{1}{iq_l-2E_{\bf k}}-\frac{1}{iq_l+2E_{\bf k}}\right)\left[1-f(E_{\bf k}-\mu)-f(E_{\bf k}+\mu)\right].
\end{eqnarray}
The coupling function is given by
\begin{eqnarray}
C^0_{\rm m}(Q)&=&\frac{1}{\beta V}\sum_{K}{\rm Tr}\left[{\cal G}(K)\gamma^0
\tau_3\gamma^5{\cal G}(K+Q)\Gamma_{\rm m}\right].
\end{eqnarray}
At ${\bf q}=0$, we can show that $C^0_{\rm m}(iq_l, {\bf q}=0)$ vanish for ${\rm m}=0,1,2$. The nonzero coupling $C^0_{\rm 3}(iq_l, {\bf q}=0)$ is given by
\begin{eqnarray}
C^0_{\rm 3}(iq_l, {\bf q}=0)=2iN_c N_f\int{d^3{\bf k}\over (2\pi)^3} \frac{M}{E_{\bf k}}
\left(\frac{1}{iq_l-2E_{\bf k}}+\frac{1}{iq_l+2E_{\bf k}}\right)\left[1-f(E_{\bf k}-\mu)-f(E_{\bf k}+\mu)\right].
\end{eqnarray}
Thus the axial isospin density response couples to the neutral pion mode $\pi_0$.

\begin{figure}
\centering
\includegraphics[width=3.05in]{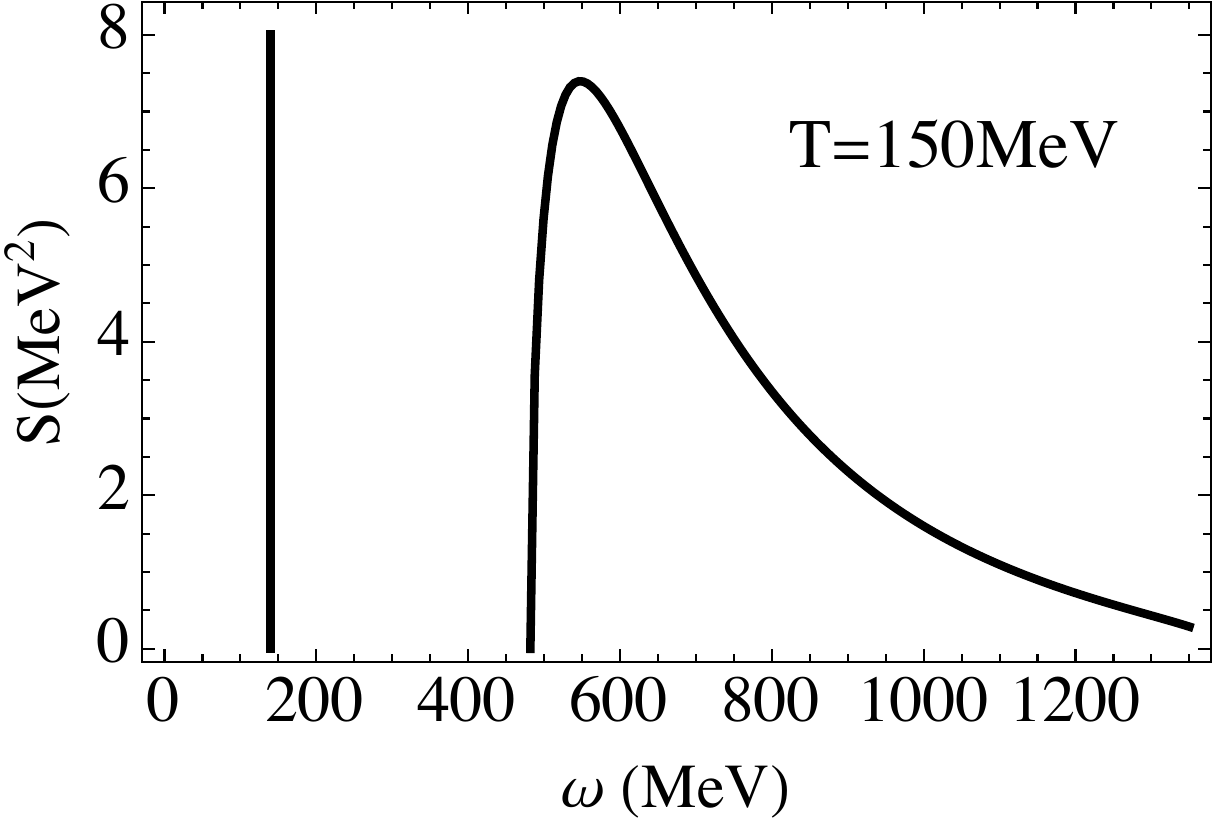}
\includegraphics[width=3.5in]{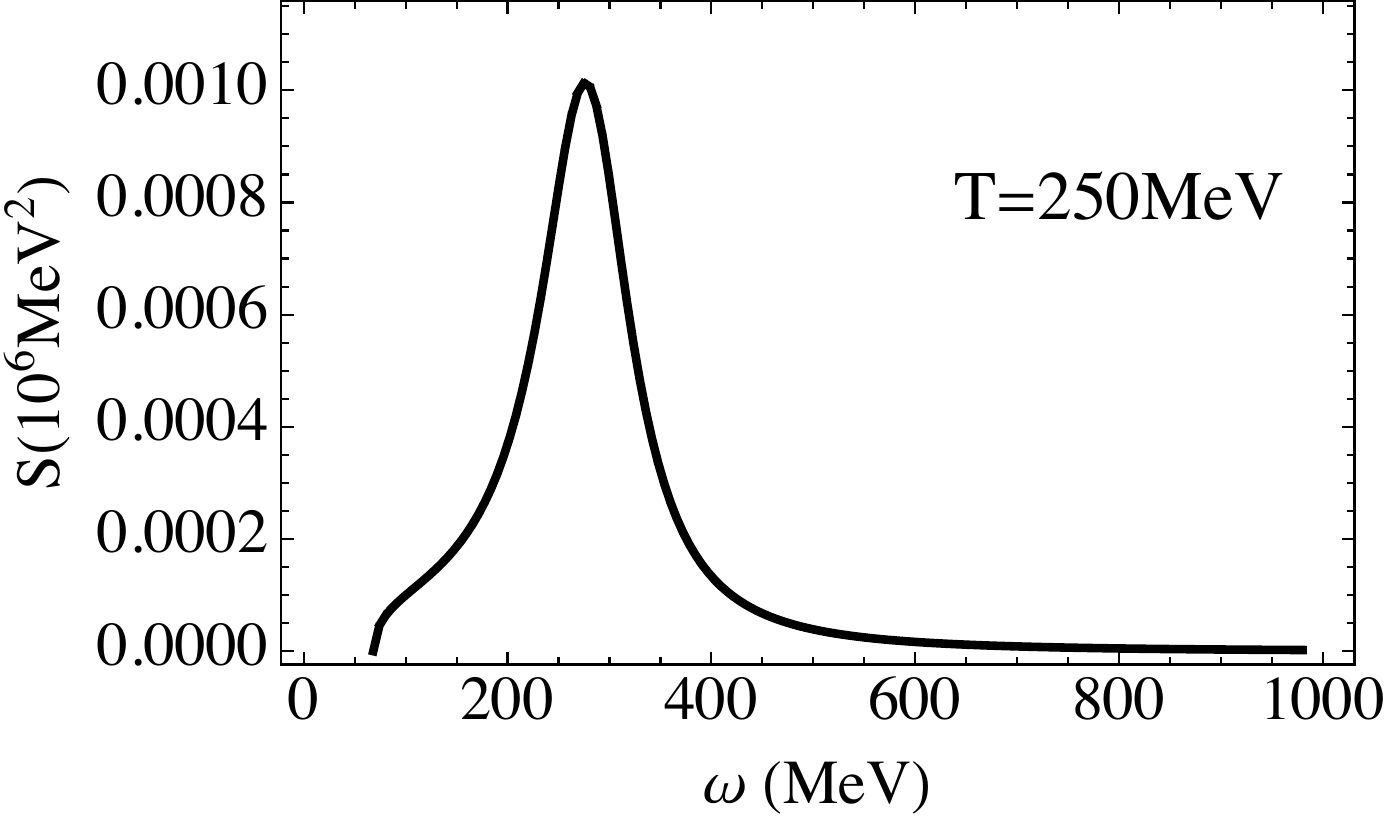}
\caption{ The dynamical structure factor $S(\omega,{\bf q})$ at ${\bf q}=0$ for the axial isospin density response ($\hat{X}=\tau_3\gamma_5$) below and above the
pion Mott transition temperature $T_{\rm M}$. We consider physical current quark mass $m_0=5$MeV and $\mu=0$. The pion Mott transition temperature in this case is
$T_{\rm M}=193$MeV.} \label{fig4}
\end{figure}

The full response function reads
\begin{eqnarray}
\chi(iq_l,{\bf q}=0)=\Pi_{\rm b}^{00}(iq_l,{\bf q}=0)-\frac{C^0_{\rm
3}(iq_l,{\bf q}=0)C^0_{\rm 3}(-iq_l,{\bf q}=0)}{\frac{1}{2G}+\Pi_{\rm 3}(iq_l,{\bf q}=0)}.
\end{eqnarray}
Here  $\Pi_{\rm 3}(iq_l,{\bf q}=0)$ is given by
\begin{eqnarray}
\Pi_{\rm 3}(iq_l,{\bf q}=0)
=2N_c N_f\int{d^3{\bf k}\over
(2\pi)^3}\left(\frac{1}{iq_l-2E_{\bf k}}-\frac{1}{iq_l+2E_{\bf k}}\right)\left[1-f(E_{\bf k}-\mu)-f(E_{\bf k}+\mu)\right].
\end{eqnarray}
To evaluate the dynamical structure factor, we make use of the following results,
\begin{eqnarray}
{\rm Im}\Pi_{\rm b}^{00}(\omega+i\epsilon,{\bf q}=0)&=&-N_c N_f \frac{M^2}{2\pi}\frac{\sqrt{\omega^2-4M^2}}{\omega}\Theta(|\omega|-2M) \left[1-\frac{1}{e^{\beta(\frac{1}{2}\omega-\mu)}+1}-\frac{1}{e^{\beta(\frac{1}{2}\omega+\mu)}+1}\right],\nonumber\\
{\rm Re}\Pi_{\rm 3}(\omega+i\epsilon,{\bf q}=0)
&=&2N_c N_f {\cal P}\int{d^3{\bf k}\over
(2\pi)^3}\left(\frac{1}{\omega-2E_{\bf k}}-\frac{1}{\omega+2E_{\bf k}}\right)\left[1-f(E_{\bf k}-\mu)-f(E_{\bf k}+\mu)\right],\nonumber\\
{\rm Im}\Pi_{\rm 3}(\omega+i\epsilon,{\bf q}=0)&=&-N_c N_f \frac{\omega\sqrt{\omega^2-4M^2}}{8\pi} \Theta(|\omega|-2M) 
\left[1-\frac{1}{e^{\beta(\frac{1}{2}\omega-\mu)}+1}-\frac{1}{e^{\beta(\frac{1}{2}\omega+\mu)}+1}\right],\nonumber\\
{\rm Re}C^0_{\rm 3}(\omega+i\epsilon,{\bf q}=0)&=&N_c N_f \frac{M\sqrt{\omega^2-4M^2}}{4 \pi} \Theta(|\omega|-2M){\rm sgn}(\omega)
\left[1-\frac{1}{e^{\beta(\frac{1}{2}\omega-\mu)}+1}-\frac{1}{e^{\beta(\frac{1}{2}\omega+\mu)}+1}\right],\nonumber\\
{\rm Im}C^0_{\rm 3}(\omega+i\epsilon, {\bf q}=0)&=&2N_c N_f{\cal P}\int{d^3{\bf k}\over (2\pi)^3} \frac{M}{E_{\bf k}}
\left(\frac{1}{\omega-2E_{\bf k}}+\frac{1}{\omega+2E_{\bf k}}\right)\left[1-f(E_{\bf k}-\mu)-f(E_{\bf k}+\mu)\right].
\end{eqnarray}
The imaginary part of $\chi(\omega+i\epsilon,{\bf q}=0)$ can be expressed as
\begin{eqnarray}
{\rm Im}\chi(\omega+i\epsilon)=\frac{{\rm Im}\Pi_{\rm 3}(\omega+i\epsilon)}{\left[\frac{1}{2G}+{\rm Re}\Pi_{\rm 3}(\omega+i\epsilon)\right]^2+\left[{\rm Im}\Pi_{\rm 3}(\omega+i\epsilon)\right]^2}
\left[{\rm Im}C^0_{\rm 3}(\omega+i\epsilon)-\frac{2M}{\omega}\left(\frac{1}{2G}+{\rm Re}\Pi_{\rm 3}(\omega+i\epsilon)\right)\right]^2.
\end{eqnarray}
Here we have suppressed the condition ${\bf q}=0$ for convenience. Therefore, we expect that at low temperature, the dynamical structure factor for the axial isospin density response reveals a pole plus continuum structure. For $|\omega|>2M$,  ${\rm Im}\Pi_{\rm 3}(\omega+i\epsilon)$ is nonzero and hence the dynamical structure factor shows a continuum. For $|\omega|<2M$, ${\rm Im}\Pi_{\rm 3}(\omega+i\epsilon)$ 
vanishes and  thus the dynamical structure factor is just proportional to a delta function. We have
\begin{eqnarray}
{\rm Im}\chi(\omega+i\epsilon)=\pi\left[{\rm Im}C^0_{\rm 3}(\omega+i\epsilon)\right]^2
\delta{\left(\frac{1}{2G}+{\rm Re}\Pi_{\rm 3}(\omega+i\epsilon)\right)}.
\end{eqnarray}
It is evident that the pole is just located at the pion mass. In the chiral limit, this pole is located exactly at $\omega=0$ for $T<T_c$ and it disappears for $T>T_c$. For physical current quark mass, therefore exists a Mott transition
temperature $T=T_{\rm M}$ determined by the equation $m_\pi (T)=2M(T)$. Figure \ref{fig4} shows the dynamical structure factor $S(\omega,{\bf q}=0)$ for temperatures below and above $T_{\rm M}$. For $T<T_{\rm M}$, the pion is a bound state and hence $S(\omega,{\bf q}=0)$ shows a pole plus continuum structure. Above the Mott transition temperature, the pole disappears and $S(\omega,{\bf q}=0)$ shows only a continuum.  The threshold of the continuum is also located at $\omega_{\rm th}=2M$.

\section{Linear Response beyond Random Phase Approximation: \\ Meson-fluctuation contribution}\label{s7}

In the chiral limit ($m_0=0$), the quarks become massless ($M=0$) above the chiral phase transition temperature. In this case we can show that $C^\mu_{\rm m}(Q)=0$ in the chiral symmetry
restored phase. Therefore, the quasi-particle random phase approximation just describes the linear response of a system of noninteracting massless quarks. However, it is generally expected that
the mesonic fluctuations play important role above and near the chiral phase transition, indicating that the random phase approximation is inadequate for such a strongly interacting system.  In this part, we 
consider a linear response theory beyond the random phase approximation.  To this end,  we recall that the generating functional in the Gaussian approximation can be expressed as
\begin{eqnarray}
{\cal W}_{\rm NJL}[A] = {\cal W}_{\rm MF}[A]+{\cal W}_{\rm FL}[A].
\end{eqnarray}
In the previous random phase approximation, the meson-fluctuation contribution ${\cal W}_{\rm FL}[A]$ is neglected.  We expect that this part becomes rather important near and above the chiral phase transition, where the quarks become massless and the mesonic degrees of freedom are still important. We note that this is a general feature of a strongly interacting fermionic system. In strong-coupling superconductors, the pair fluctuation has important
contribution to the transport properties above and near the superconducting transition temperature \cite{AL1968,MT1968}.

Now we consider the contribution from the mesonic fluctuations.  To derive the generating
functional ${\cal W}_{\rm FL}[A]$, we first note that
\begin{eqnarray}
{\bf G}_{\rm A}^{-1}(x,x^\prime)={\cal G}_{\rm A}^{-1}(x,x^\prime)-\Sigma_{\rm
FL}(x,x^\prime),
\end{eqnarray}
where $\Sigma_{\rm FL}$ includes the mesonic fluctuation fields,
\begin{eqnarray}
\Sigma_{\rm FL}(x,x^\prime)=\sum_{\rm m=0}^3\Gamma_{\rm m}\phi_{\rm
m}(x)\delta(x-x^\prime),
\end{eqnarray}
and ${\cal G}_{\rm A}^{-1}(x,x^\prime)$ is the mean-field quark Green's function with external source,
\begin{eqnarray}
{\cal G}_{\rm A}^{-1}(x,x^\prime)={\cal G}^{-1}(x,x^\prime)-\Sigma_{\rm A}(x,x^\prime),
\end{eqnarray}
with ${\cal G}^{-1}(x,x^\prime)$ and $\Sigma_{\rm A}(x,x^\prime)$ given in Eq. (79). Converting to the momentum space, we have
\begin{eqnarray}
&&{\bf G}_{\rm A}^{-1}(K,K^\prime)={\cal G}_{\rm A}^{-1}(K,K^\prime)-\Sigma_{\rm FL}(K,K^\prime)\nonumber\\
&&\Sigma_{\rm FL}(K,K^\prime)=\sum_{\rm m=0}^3\Gamma_{\rm m}\phi_{\rm m}(K-K^\prime).
\end{eqnarray}
Starting from Eqs. (67) and (68) and applying the derivative expansion, we obtain
\begin{eqnarray}
{\cal S}_{\rm{eff}}^{(2)}[\sigma,\mbox{\boldmath{$\pi$}}]=\frac{\beta V}{2}\sum_{\rm
m,n=0}^3\sum_{Q,Q^\prime}\phi_{\rm m}(-Q)\left[{\bf D}_{\rm
A}^{-1}(Q,Q^\prime)\right]_{\rm mn}\phi_{\rm n}(Q^\prime),
\end{eqnarray}
where
\begin{eqnarray}
\left[{\bf D}_{\rm A}^{-1}(Q,Q^\prime)\right]_{\rm mn}=\frac{\delta_{\rm
mn}}{2G}\delta_{Q,Q^\prime}+\frac{1}{\beta V}\sum_{K,K^\prime} {\rm Tr}\left[{\cal
G}_{\rm A}(K,K^\prime-Q)\Gamma_{\rm m}{\cal G}_{\rm A}(K^\prime,K+Q^\prime)\Gamma_{\rm
n}\right].
\end{eqnarray}
Again, the path integral over the fluctuation fields $\phi_{\rm m}$ can be worked out
and we obtain
\begin{eqnarray}
{\cal W}_{\rm FL}[A]=\frac{1}{2}{\rm Tr}\ln[{\bf D}_{\rm A}^{-1}(Q,Q^\prime)].
\end{eqnarray}
Note that the trace here is also taken in the momentum space.

The next step is to expand ${\cal W}_{\rm FL}[A]$ in powers the external
source and the induced perturbations. To this end, we first need to expand the inverse meson propagator ${\bf D}_{\rm A}^{-1}(Q,Q^\prime)$ in powers of
$A_\mu$ and $\eta_{\rm m}$. The expansion takes the form
\begin{equation}
{\bf D}_{\rm A}^{-1}(Q,Q^\prime)={\bf
D}^{-1}(Q)\delta_{Q,Q^\prime}+\Sigma^{(1)}(Q,Q^\prime)+\Sigma^{(2)}(Q,Q^\prime)+\cdots.
\end{equation}
Here ${\bf D}(Q)$ is the meson propagator evaluated in Sec. \ref{s3}, and $\Sigma^{(n)}$ denotes the $n$th-order expansion in $A_\mu$ and $\eta_{\rm m}$. In practice, we only need to evaluate the expansion
up to the second order, since the higher order contributions are irrelevant to the linear response.  Like ${\bf D}^{-1}$, $\Sigma^{(1)}$ and $\Sigma^{(2)}$ are $4\times4$
matrices in the space spanned by ${\rm m}=0,1,2,3$.

To obtain $\Sigma^{(1)}$ and $\Sigma^{(2)}$, we note that in the momentum space, the
inverse of the mean-field quark Green's function with external source, ${\cal G}_{\rm A}^{-1}$,  is given by
\begin{equation}
{\cal G}_{\rm A}^{-1}(K,K^\prime)={\cal G}^{-1}(K)\delta_{K,K^\prime}-\Sigma_{\rm
A}(K,K^\prime),
\end{equation}
where
\begin{eqnarray}
\Sigma_{\rm A}(K,K^\prime)=\sum_{\rm m=0}^3\Gamma_{\rm m}\eta_{\rm
m}(K-K^\prime)+\Gamma^\mu A_\mu(K-K^\prime).
\end{eqnarray}
For convenience, here we express $\Sigma_{\rm A}$ in a more compact form
\begin{eqnarray}
\Sigma_{\rm A}(K,K^\prime)=\sum_{\rm i=0}^7\tilde{\Gamma}^{\rm i}\Phi_{\rm i}(K-K^\prime),
\end{eqnarray}
where $\tilde{\Gamma}$ is a compact notation of $(\Gamma_{\rm m},\Gamma^\mu)$ and $\Phi$ is a compact form of $(\eta_{\rm m},A_\mu)$.
Here ${\rm i}=0,1,2,3$ still stands for $\eta_{\rm m}$ with ${\rm m}=0,1,2,3$, and ${\rm i}=4,5,6,7$ stands for $A_\mu$ with $\mu=0,1,2,3$. Applying the Taylor expansion for matrix functions, we obtain
\begin{equation}
{\cal G}_{\rm A}={\cal G}+{\cal G}\Sigma_{\rm A}{\cal G}+{\cal G}\Sigma_{\rm A}{\cal
G}\Sigma_{\rm A}{\cal G}+\cdots.
\end{equation}
This compact form of the Taylor expansion should be understood in all spaces. In the momentum space, we have explicitly
\begin{eqnarray}
{\cal G}_{\rm A}(K,K^\prime)&=&{\cal G}(K,K^\prime)+\sum_{K_1,K_2}{\cal G}(K,K_1)\Sigma_{\rm A}(K_1,K_2){\cal G}(K_2,K^\prime)\nonumber\\
&+&\sum_{K_1,K_2,K_3,K_4}{\cal G}(K,K_1)\Sigma_{\rm A}(K_1,K_2){\cal
G}(K_2,K_3)\Sigma_{\rm A}(K_3,K_4){\cal G}(K_4,K^\prime)+\cdots.
\end{eqnarray}
According to the fact that ${\cal G}(K,K^\prime)={\cal G}(K)\delta_{K,K^\prime}$ and
$\Sigma_{\rm A}(K,K^\prime)=\Sigma_{\rm A}(K-K^\prime)$, it can be simplified as
\begin{eqnarray}
{\cal G}_{\rm A}(K,K^\prime)&=&{\cal G}(K)\delta_{K,K^\prime}+{\cal G}(K)\Sigma_{\rm A}(K-K^\prime){\cal G}(K^\prime)\nonumber\\
&+&\sum_{K^{\prime\prime}}{\cal G}(K)\Sigma_{\rm A}(K-K^{\prime\prime}){\cal
G}(K^{\prime\prime})\Sigma_{\rm A}(K^{\prime\prime}-K^\prime){\cal
G}(K^{\prime})+\cdots.
\end{eqnarray}

The explicit form of $\Sigma^{(1)}$ and $\Sigma^{(2)}$ can be derived by using the above expansion for ${\cal G}_{\rm A}$. $\Sigma^{(1)}$ is composed of one zeroth-order 
and one first-order contributions of ${\cal G}_{\rm A}$. It is explicitly given by
\begin{eqnarray}
\Sigma^{(1)}_{\rm mn}(Q,Q^\prime)=\sum_{\rm i=0}^7{[X^{(1)}]}^{\rm i}_{\rm mn}(Q,Q^\prime)\Phi_{\rm i}(Q-Q^\prime),
\end{eqnarray}
where the coefficients are given by
\begin{eqnarray}
{[X^{(1)}]}^{\rm i}_{\rm mn}(Q,Q^\prime)&=&\frac{1}{\beta V}\sum_K{\rm Tr}\left[{\cal
G}(K)\Gamma_{\rm m}{\cal G}(K+Q)\tilde{\Gamma}^{\rm i}
{\cal G}(K+Q^\prime)\Gamma_{\rm n}\right]\nonumber\\
&+&\frac{1}{\beta V}\sum_K{\rm Tr}\left[{\cal G}(K)\tilde{\Gamma}^{\rm i}{\cal G}(K+Q^\prime-Q)\Gamma_{\rm m}{\cal G}(K+Q^\prime)\Gamma_{\rm n}\right].
\end{eqnarray}
$\Sigma^{(2)}$ includes two types of contributions. We have
\begin{eqnarray}
\Sigma^{(2)}=\Sigma^{(2a)}+\Sigma^{(2b)}.
\end{eqnarray}
$\Sigma^{(2a)}$ is composed of one zeroth-order and one second-order contributions of ${\cal G}_{\rm A}$. It is given by
\begin{eqnarray}
\Sigma^{(2a)}_{\rm mn}(Q,Q^\prime)&=&\frac{1}{\beta V}\sum_{{\rm i,j}=0}^7\sum_{K,K^\prime}{[X^{(2a)}]}_{\rm mn}^{\rm ij}(Q,Q^\prime;K,K^\prime)\Phi_{\rm i}(Q_1)\Phi_{\rm j}(Q_2)\nonumber\\
&+&\frac{1}{\beta V}\sum_{{\rm i,j}=0}^7\sum_{K,K^\prime}{[Y^{(2a)}]}_{\rm mn}^{\rm ij}(Q,Q^\prime;K,K^\prime)\Phi_{\rm i}(Q_3)\Phi_{\rm j}(Q_4).
\end{eqnarray}
Here the momenta $Q_1,Q_2,Q_3$, and $Q_4$ are defined as
\begin{eqnarray}
&&Q_1=K-K^\prime+Q,\ \ \ \ \ \ Q_2=K^\prime-K-Q^\prime,\nonumber\\
&&Q_3=K-K^\prime,\ \ \ \ \ \ Q_4=K^\prime-K+Q-Q^\prime.
\end{eqnarray}
The expansion coefficients are given by
\begin{eqnarray}
&&{[X^{(2a)}]}_{\rm mn}^{\rm ij}(Q,Q^\prime;K,K^\prime)={\rm Tr}\left[{\cal G}(K)\Gamma_{\rm m}{\cal G}(K+Q)\tilde{\Gamma}^{\rm i}{\cal G}(K^\prime)\tilde{\Gamma}^{\rm j}{\cal G}(K+Q^\prime)\Gamma_{\rm n}\right],
\nonumber\\
&&{[Y^{(2a)}]}_{\rm mn}^{\rm ij}(Q,Q^\prime;K,K^\prime)={\rm Tr}\left[{\cal G}(K)\tilde{\Gamma}^{\rm i}{\cal G}(K^\prime)\tilde{\Gamma}^{\rm j}{\cal G}(K+Q^\prime-Q) 
\Gamma_{\rm m}{\cal G}(K+Q^\prime)\Gamma_{\rm n}\right].
\end{eqnarray}
$\Sigma^{(2b)}$ is composed of two second-order contributions of ${\cal G}_{\rm A}$. It reads
\begin{eqnarray}
\Sigma^{(2b)}_{\rm mn}(Q,Q^\prime)=\frac{1}{\beta V}\sum_{{\rm i,j}=0}^7\sum_{K,K^\prime}{[X^{(2b)}]}_{\rm mn}^{\rm ij}(Q,Q^\prime;K,K^\prime)\Phi_{\rm i}(Q_1)\Phi_{\rm j}(Q_2),
\end{eqnarray}
where the expansion coefficient is given by
\begin{eqnarray}
{[X^{(2b)}]}_{\rm mn}^{\rm ij}(Q,Q^\prime;K,K^\prime)={\rm Tr}\left[{\cal G}(K)\tilde{\Gamma}^{\rm i}{\cal
G}(K^\prime-Q)\Gamma_{\rm m}{\cal G}(K^\prime)\tilde{\Gamma}^{\rm j}{\cal G}(K+Q^\prime)\Gamma_{\rm
n}\right].
\end{eqnarray}

\subsection{Derivation of various contributions}

Now we express the meson-fluctuation contribution to the generating functional as
\begin{eqnarray}
{\cal W}_{\rm FL}[A;\eta]=\frac{1}{2}{\rm Tr}\ln\left[{\bf
D}^{-1}(Q)\delta_{Q,Q^\prime}+\Sigma^{(1)}(Q,Q^\prime)+\Sigma^{(2a)}(Q,Q^\prime)
+\Sigma^{(2b)}(Q,Q^\prime)+\cdots\right].
\end{eqnarray}
From now we show the explicit dependence on the induced perturbations. Applying the trick of derivative expansion, we can expand ${\cal W}_{\rm FL}$ in powers of 
the external source as well as the induced perturbations. We have
\begin{eqnarray}
{\cal W}_{\rm FL}[A;\eta]={\cal W}_{\rm FL}^{(0)}+{\cal W}_{\rm FL}^{(1)}[A;\eta]+{\cal
W}_{\rm FL}^{(2)}[A;\eta]+\cdots.
\end{eqnarray}
It is evident that ${\cal W}_{\rm FL}^{(0)}=\beta V\Omega_{\rm FL}$ and hence the present linear response theory including the meson-fluctuation contribution is parallel to the meson-fluctuation theory of the 
equilibrium thermodynamics. The first-order expansion is given by
\begin{eqnarray}\label{W-linear}
{\cal W}_{\rm FL}^{(1)}[A;\eta]=\frac{1}{2}\sum_{Q}{\rm Tr}_{4\rm D}\left[{\bf
D}(Q)\Sigma^{(1)}(Q,Q)\right].
\end{eqnarray}
Here the trace ${\rm Tr}_{4\rm D}$ is now taken only in the four-dimensional space
spanned by ${\rm m},{\rm n}=0,1,2,3$. The second-order expansion can be expressed as
\begin{eqnarray}
{\cal W}_{\rm FL}^{(2)}[A;\eta]={\cal W}_{\rm FL}^{({\rm AL})}+{\cal W}_{\rm FL}^{({\rm
SE})}+{\cal W}_{\rm FL}^{({\rm MT})},
\end{eqnarray}
where the three kinds of contributions are given by
\begin{eqnarray}
&&{\cal W}_{\rm FL}^{({\rm AL})}[A;\eta]=-\frac{1}{4}\sum_{Q}\sum_{Q^\prime}
{\rm Tr}_{4\rm D}\left[{\bf D}(Q)\Sigma^{(1)}(Q,Q^\prime){\bf D}(Q^\prime)\Sigma^{(1)}(Q^\prime,Q)\right],\nonumber\\
&&{\cal W}_{\rm FL}^{({\rm SE})}[A;\eta]=\frac{1}{2}\sum_{Q}{\rm Tr}_{4\rm D}\left[{\bf D}(Q)\Sigma^{(2a)}(Q,Q)\right],\nonumber\\
&&{\cal W}_{\rm FL}^{({\rm MT})}[A;\eta]=\frac{1}{2}\sum_{Q}{\rm Tr}_{4\rm D}\left[{\bf
D}(Q)\Sigma^{(2b)}(Q,Q)\right],
\end{eqnarray}
which correspond diagrammatically to the Aslamazov-Lakin (AL), Self-Energy (SE) or Density-of-State, and Maki-Thompson (MT) contributions.

To obtain the response functions, we need to eliminate the induced perturbations $\eta_{\rm m}(Q)$. Noting that the present theory of linear response is a natural generalization of the meson-fluctuation theory of the 
equilibrium thermodynamics, where the order parameter is determined at the mean-field level, we determine the induced perturbations $\eta_{\rm m}(Q)$ still by minimizing the mean-field generation functional, i.e.,
\begin{equation}
\frac{\delta{\cal W}_{\rm MF}^{(2)}[A;\eta]}{\delta \eta_{\rm m}(Q)}=0,
\end{equation}
which leads to
\begin{equation}
\eta_{\rm m}(Q)=-\frac{C^\mu_{\rm m}(-Q)}{\frac{1}{2G}+\Pi_{\rm m}(Q)}A_\mu(Q)+O(A^2),\
\ \ \ \ \ \eta_{\rm m}(-Q)=-\frac{C^\mu_{\rm m}(Q)}{\frac{1}{2G}+\Pi_{\rm
m}(Q)}A_\mu(-Q)+O(A^2).
\end{equation}
Later we will show that the use of the above relations is also crucial to recover the correct number susceptibility in the static and long-wavelength limit.

\subsubsection{Order parameter induced contribution}
Unlike the mean-field theory or random phase approximation, the first-order contribution, Eq. (\ref{W-linear}), becomes highly nontrivial. It can be
expressed as
\begin{eqnarray}\label{W-linear2}
{\cal W}_{\rm FL}^{(1)}[A;\eta]=\beta V\sum_{{\rm i}=1}^7{\cal C}_{\rm i}\Phi_{\rm i}(0),
\end{eqnarray}
where the coefficients read
\begin{eqnarray}
{\cal C}_{\rm i}=\frac{1}{2\beta V}\sum_{Q}{\rm Tr}_{4\rm D}\left\{{\bf D}(Q)[X^{(1)}]^{\rm i}(Q,Q)\right\}.
\end{eqnarray}
Using the explicit expression of $X^{(1)}$, we can show that possible nonvanishing coefficients are
\begin{equation}
{\cal C}_0=\frac{\partial \Omega_{\rm FL}(M,\mu_{\rm X})}{\partial M}, \ \ \ \ \ \ \ \ \ \ \ \  {\cal C}_4=\frac{\partial \Omega_{\rm FL}(M,\mu_{\rm X})}{\partial\mu_{\rm X}}.
\end{equation}
Since we consider only nonzero baryon chemical potential, here the effective chemical potential $\mu_{\rm X}=A_0(0)$ is nonvanishing only for the vector current case $(\hat{X}=1)$.  Thus ${\cal C}_4$ is nonvanishing only for the case $\hat{X}=1$, where $\mu_{\rm X}$ corresponds to the quark chemical potential $\mu$. The fact ${\cal C}_0\neq0$ indicates that the first-order contribution
${\cal W}_{\rm FL}^{(1)}[A;\eta]$ cannot be simply neglected, since it does contributes to the linear response.  To understand this, we note that when eliminating the
induced perturbation $\eta_0(0)$, Eq. (147) is not adequate. Actually, the contributions of the order $O(A^2)$ in Eq. (147) becomes important. To obtain these contributions, we should expand the mean-field  generating functional  ${\cal W}_{\rm MF}[A;\eta]$ up to the third order in $A$ and $\eta$. We have
\begin{eqnarray}
{\cal W}_{\rm MF}^{(3)}[A;\eta]=\frac{1}{3}\sum_{K}\sum_{K^\prime}\sum_{K^{\prime\prime}}{\rm
Tr}\left[{\cal G}(K)\Sigma_{\rm A}(K,K^\prime){\cal G}(K^\prime)
\Sigma_{\rm A}(K^\prime,K^{\prime\prime})
{\cal G}(K^{\prime\prime})\Sigma_{\rm A}(K^{\prime\prime},K)\right].
\end{eqnarray}
By defining $K^\prime=K+Q$ and $K^{\prime\prime}=K+Q^\prime$, we obtain
\begin{eqnarray}
\frac{{\cal W}_{\rm MF}^{(3)}[A;\eta]}{\beta V}=\frac{1}{3}\sum_{{\rm i,j,k}=0}^7\sum_{Q}\sum_{Q^\prime}F_{\rm ijk}(Q,Q^\prime)
\Phi_{\rm i}(-Q)\Phi_{\rm j}(Q-Q^\prime)\Phi_{\rm k}(Q^\prime),
\end{eqnarray}
where the function $F_{\rm ijk}(Q,Q^\prime)$ is defined as
\begin{eqnarray}
F_{\rm ijk}(Q,Q^\prime)=\frac{1}{\beta V}\sum_{K}{\rm Tr}\left[{\cal
G}(K)\tilde{\Gamma}^{\rm i}{\cal G}(K+Q)\tilde{\Gamma}^{\rm j}{\cal G}(K+Q^\prime)\tilde{\Gamma}^{\rm k}\right],
\end{eqnarray}
Using the extreme condition
\begin{eqnarray}
\frac{\delta{\cal W}_{\rm MF}[A;\eta]}{\delta\eta_{0}(Q)}=0
\end{eqnarray}
with ${\cal W}_{\rm MF}={\cal W}_{\rm MF}^{(0)}+{\cal W}_{\rm MF}^{(1)}+{\cal W}_{\rm
MF}^{(2)}+{\cal W}_{\rm MF}^{(3)}+\cdots$, we obtain
\begin{eqnarray}\label{Exp-eta}
\eta_0(0)={\cal R}_1A_0(0)+\frac{1}{2}\sum_{{\rm i,j}=0}^7
\sum_Q{\cal U}_{\rm ij}(Q)\Phi_{\rm i}(-Q)\Phi_{\rm j}(Q)+\cdots,
\end{eqnarray}
where the coefficients ${\cal R}_1$ and ${\cal U}_{\rm ij}(Q)$ are given by
\begin{eqnarray}
&&{\cal R}_1=-\lim_{Q\rightarrow0}\frac{C^0_0(-Q)}{\frac{1}{2G}+\Pi_0(Q)},\nonumber\\
&&{\cal U}_{\rm ij}(Q)=\frac{2}{3}\lim_{Q^\prime\rightarrow0}\frac{F_{0{\rm ij}}(-Q^\prime,Q)+F_{{\rm i}0{\rm j}}(Q+Q^\prime,Q)+F_{{\rm ij}0}(Q,Q^\prime)}{\frac{1}{2G}+\Pi_0(Q)}.
\end{eqnarray}
Here the the static and long-wavelength limit of an arbitrary function ${\cal A}(Q)$ should be understood as
$\lim_{Q\rightarrow0}{\cal A}(Q)=\lim_{{\bf q}\rightarrow0}{\cal A}(iq_l=0,{\bf q})$.
For the purpose of linear response we apply Eq. (147) and obtain
\begin{eqnarray}\label{Exp-eta}
\eta_0(0)={\cal R}_1A_0(0)+\frac{1}{2}
\sum_Q{\cal R}_2^{\mu\nu}(Q)A_{\mu}(-Q)A_{\nu}(Q)+O(A^3),
\end{eqnarray}
where the explicit form of the function ${\cal R}_2^{\mu\nu}(Q)$ is not shown here. It is evident that
\begin{equation}
{\cal R}_1=\frac{\partial M(\mu_{\rm X})}{\partial \mu_{\rm X}},  \ \ \ \ \ \ \lim_{Q\rightarrow0}{\cal R}_2^{00}(Q)=\frac{\partial^2 M(\mu_{\rm X})}{\partial \mu_{\rm X}^2}.
\end{equation}

Substituting the expansion (\ref{Exp-eta}) into Eq. (\ref{W-linear2}), we eliminate the induced perturbations and obtain
\begin{eqnarray}
\frac{{\cal W}_{\rm FL}^{(1)}[A]}{\beta V}=- (n_{\rm X})_{\rm
FL}A_0(0)+\frac{1}{2}
\sum_Q\Pi_{\rm OP}^{\mu\nu}(Q)A_{\mu}(-Q)A_{\nu}(Q)+\cdots,
\end{eqnarray}
where $(n_{\rm X})_{\rm FL}$ is the fluctuation contribution to the charge density, 
\begin{equation}
(n_{\rm X})_{\rm FL}=-\frac{\partial \Omega_{\rm FL}(M,\mu_{\rm X})}{\partial\mu_{\rm X}}-\frac{\partial \Omega_{\rm FL}(M,\mu_{\rm X})}{\partial M}\frac{\partial M(\mu_{\rm X})}{\partial \mu_{\rm X}}.
\end{equation}
The first-order term ${\cal W}_{\rm FL}^{(1)}$ thus gives a nontrivial contribution to the response function, which is given by
\begin{eqnarray}
\Pi_{\rm OP}^{\mu\nu}(Q)={\cal C}_0{\cal R}_2^{\mu\nu}(Q).
\end{eqnarray}
It is evident that this contribution is due to the nonvanishing chiral condensate. In the chiral limit, this contribution vanishes above the phase transition temperature, where ${\cal C}_0=0$.
Therefore, we denote it as the order parameter induced (OP) contribution, which can be expressed as
\begin{eqnarray}
{\cal W}_{\rm FL}^{({\rm OP})}[A]=\frac{\beta V}{2}\sum_Q\Pi_{\rm OP}^{\mu\nu}(Q)A_{\mu}(-Q)A_{\nu}(Q).
\end{eqnarray}

\subsubsection{Aslamazov-Lakin contribution}
The Aslamazov-Lakin contribution  is given by
\begin{eqnarray}
{\cal W}_{\rm FL}^{({\rm AL})}[A;\eta]=-\frac{1}{4}\sum_{Q} \sum_{Q^\prime}{\rm Tr}_{\rm 4D}
\left[{\bf D}(Q)\Sigma^{(1)}(Q,Q^\prime){\bf D}(Q^\prime)\Sigma^{(1)}(Q^\prime,Q)\right].
\end{eqnarray}
After some manipulations, it can be expressed as
\begin{eqnarray}
{\cal W}_{\rm FL}^{({\rm AL})}[A;\eta]=\frac{\beta V}{2}\sum_{{\rm i,j}=0}^7\sum_{Q}
\Xi_{\rm ij}^{\rm AL}(Q)\Phi_{\rm i}(-Q)\Phi_{\rm j}(Q),
\end{eqnarray}
where the function $\Xi_{\rm ij}^{\rm AL}(Q)$ is given by
\begin{eqnarray}
\Xi_{\rm ij}^{\rm AL}(Q)=-\frac{1}{2}\frac{1}{\beta V}\sum_{P}{\rm Tr}_{\rm 4D}
\left\{{\bf D}(P)[X^{(1)}]^{\rm i}(P,P+Q){\bf D}(P+Q)[X^{(1)}]^{\rm j}(P+Q,P)\right\}.
\end{eqnarray}
Here the matrices $[X^{(1)}]^{\rm i}(P,P+Q)$ and $[X^{(1)}]^{\rm j}(P+Q,P)$ are defined as
\begin{eqnarray}
[X^{(1)}]^{\rm i}_{\rm mn}(P,P+Q)&=&\frac{1}{\beta V}\sum_K{\rm Tr}
\left[{\cal G}(K)\Gamma_{\rm m}{\cal G}(K+P)\tilde{\Gamma}^{\rm i}
{\cal G}(K+P+Q)\Gamma_{\rm n}\right]\nonumber\\
&+&\frac{1}{\beta V}\sum_K{\rm Tr}\left[{\cal G}(K)\tilde{\Gamma}^{\rm i}{\cal G}(K+Q)\Gamma_{\rm m}{\cal G}(K+P+Q)\Gamma_{\rm n}\right],\nonumber\\
{[X^{(1)}]}^{\rm j}_{\rm mn}(P+Q,P)&=&\frac{1}{\beta V}\sum_K{\rm Tr}\left[{\cal
G}(K)\Gamma_{\rm m}{\cal G}(K+P+Q)\tilde{\Gamma}^{\rm j}
{\cal G}(K+P)\Gamma_{\rm n}\right]\nonumber\\
&+&\frac{1}{\beta V}\sum_K{\rm Tr}\left[{\cal G}(K)\tilde{\Gamma}^{\rm j}{\cal
G}(K-Q)\Gamma_{\rm m}{\cal G}(K+P)\Gamma_{\rm n}\right].
\end{eqnarray}
We finally use Eq. (147) to eliminate the induced perturbations and obtain the AL contribution
\begin{eqnarray}
{\cal W}_{\rm FL}^{({\rm AL})}[A]=\frac{\beta V}{2}\sum_{Q}
\Pi_{\rm AL}^{\mu\nu}(Q)A_{\mu}(-Q)A_{\nu}(Q),
\end{eqnarray}
where $\Pi_{\rm AL}^{\mu\nu}(Q)$ is the AL contribution to the response function.

\subsubsection{Self-Energy contribution}
The Self-Energy or Density-of-State contribution is given by
\begin{eqnarray}
{\cal W}_{\rm FL}^{({\rm SE})}[A;\eta]=\frac{1}{2}\sum_{Q}{\rm Tr}_{\rm 4D}\left[{\bf D}(Q)\Sigma^{(2a)}(Q,Q)\right],
\end{eqnarray}
After some manipulations, it can be expressed as
\begin{eqnarray}
{\cal W}_{\rm FL}^{({\rm SE})}[A;\eta]=\frac{\beta V}{2}\sum_{{\rm i,j}=1}^8\sum_{Q}\Xi_{\rm ij}^{\rm SE}(Q)\Phi_{\rm i}(-Q)\Phi_{\rm j}(Q),
\end{eqnarray}
where the function $\Xi_{\rm ij}^{\rm SE}(Q)$ is given by
\begin{eqnarray}
\Xi_{\rm ij}^{\rm SE}(Q)=\frac{1}{\beta V}\sum_{P}{\rm Tr}_{4\rm D}\left[{\bf D}(P){\bf Y}^{\rm ij}(P,Q)\right]
+\frac{1}{\beta V}\sum_{P}{\rm Tr}_{\rm 4D}\left[{\bf D}(P){\bf Z}^{\rm ij}(P,Q)\right].
\end{eqnarray}
Here the matrices ${\bf Y}^{\rm ij}(P,Q)$ and ${\bf Z}^{\rm ij}(P,Q)$ are defined as
\begin{eqnarray}
{\bf Y}_{\rm mn}^{\rm ij}(P,Q)&=&\frac{1}{\beta V}\sum_{K}{\rm Tr}\left[{\cal G}(K-P)\Gamma_{\rm m}{\cal G}(K)\tilde{\Gamma}^{\rm i}{\cal G}(K+Q)\tilde{\Gamma}^{\rm j}{\cal G}(K)\Gamma_{\rm n}\right],\nonumber\\
{\bf Z}_{\rm mn}^{\rm ij}(P,Q)&=&\frac{1}{\beta V}\sum_{K}{\rm Tr}\left[{\cal G}(K)\tilde{\Gamma}^{\rm i}{\cal G}(K+Q)\tilde{\Gamma}^{\rm j}{\cal G}(K) \Gamma_{\rm m}{\cal G}(K+P)\Gamma_{\rm n}\right].
\end{eqnarray}
We finally use Eq. (147) to eliminate the induced perturbations and obtain the SE contribution
\begin{eqnarray}
{\cal W}_{\rm FL}^{({\rm SE})}[A]=\frac{\beta V}{2}\sum_{Q}
\Pi_{\rm SE}^{\mu\nu}(Q)A_{\mu}(-Q)A_{\nu}(Q),
\end{eqnarray}
where $\Pi_{\rm SE}^{\mu\nu}(Q)$ is the SE contribution to the response function.

\subsubsection{Maki-Thompson contribution}
The Maki-Thompson contribution is given by
\begin{eqnarray}
{\cal W}_{\rm FL}^{({\rm MT})}[A;\eta]=\frac{1}{2}\sum_{Q}{\rm Tr}_{4\rm D}\left[{\bf D}(Q)\Sigma^{(2b)}(Q,Q)\right].
\end{eqnarray}
After some manipulations, it can be expressed as
\begin{eqnarray}
{\cal W}_{\rm FL}^{({\rm MT})}[A;\eta]=\frac{1}{2}\sum_{{\rm i,j}=1}^8\sum_{Q}\Xi_{\rm ij}^{\rm MT}(Q)\Phi_{\rm i}(-Q)\Phi_{\rm j}(Q),
\end{eqnarray}
where the function $\Xi_{\rm ij}^{\rm MT}(Q)$ is given by
\begin{eqnarray}
\Xi_{\rm ij}^{\rm MT}(Q)=\frac{1}{\beta V}\sum_{P}{\rm Tr}_{4\rm D}\left[{\bf D}(P){\bf W}^{\rm ij}(P,Q)\right].
\end{eqnarray}
Here the matrix ${\bf W}^{\rm ij}(P,Q)$ is defined as
\begin{eqnarray}
{\bf W}_{\rm mn}^{\rm ij}(P,Q)=\frac{1}{\beta V}\sum_{K}{\rm Tr}
\left[{\cal G}(K)\tilde{\Gamma}^{\rm i}{\cal G}(K+Q)\Gamma_{\rm m}{\cal G}(K+P+Q)\tilde{\Gamma}^{\rm j}{\cal G}(K+P)\Gamma_{\rm n}\right].
\end{eqnarray}
We finally use Eq. (147) to eliminate the induced perturbations and obtain the MT contribution
\begin{eqnarray}
{\cal W}_{\rm FL}^{({\rm MT})}[A]=\frac{\beta V}{2}\sum_{Q}
\Pi_{\rm MT}^{\mu\nu}(Q)A_{\mu}(-Q)A_{\nu}(Q),
\end{eqnarray}
where $\Pi_{\rm MT}^{\mu\nu}(Q)$ is the MT contribution to the response function.

Combining all contributions, the meson-fluctuation contribution to the linear response is given by 
\begin{eqnarray}
{\cal W}_{\rm FL}^{(2)}[A]=\frac{1}{2}\sum_{Q}\Pi_{\rm FL}^{\mu\nu}(Q)A_{\mu}(-Q)A_{\nu}(Q),
\end{eqnarray}
where $\Pi_{\rm FL}^{\mu\nu}(Q)$ is a summation of all the above contributions,
\begin{eqnarray}
\Pi_{\rm FL}^{\mu\nu}(Q)=\Pi_{\rm OP}^{\mu\nu}(Q)+\Pi_{\rm AL}^{\mu\nu}(Q)+\Pi_{\rm SE}^{\mu\nu}(Q)+\Pi_{\rm MT}^{\mu\nu}(Q).
\end{eqnarray}
Summarizing the mean-field and the meson-fluctuation contributions, we have
\begin{eqnarray}
{\cal W}_{\rm NJL}^{(2)}[A]=\frac{1}{2}\sum_{Q}\Pi^{\mu\nu}(Q)A_{\mu}(-Q)A_{\nu}(Q),
\end{eqnarray}
where the full response function within the present theory is given by
\begin{eqnarray}
\Pi^{\mu\nu}(Q)=\Pi_{\rm MF}^{\mu\nu}(Q)+\Pi_{\rm FL}^{\mu\nu}(Q).
\end{eqnarray}

\subsection{Static and long-wavelength limit}
Now we check the static and long-wavelength limit of the above linear response theory. In this limit, it is obvious that the density response function $\Pi^{00}(Q\rightarrow0)$ should recover the charge susceptibility $\kappa_{\rm X}$ associated with the
channel $X$, i.e.,
\begin{eqnarray}
\Pi^{00}(Q\rightarrow0)=-\kappa_{\rm X}=\frac{\partial^2\Omega(T,\mu_{\rm X})}{\partial \mu_{\rm X}^2}.
\end{eqnarray}
In condensed matter theory, this is the so called compressibility sum rule \cite{Levin2013}.  Here we emphasize that the correct static and long-wavelength limit of an arbitrary function ${\cal A}(Q)$ should be understood as
\begin{eqnarray}
{\cal A}(Q\rightarrow0)=\lim_{{\bf q}\rightarrow0}{\cal A}(iq_l=0,{\bf q}).
\end{eqnarray}

In the mean-field theory, the thermodynamic potential is given by $\Omega_{\rm MF}(\mu_{\rm X},M)$, where the dependence on the temperature is not explicitly shown. Note that the effective quark mass $M$ is also an implicit
function of $\mu_{\rm X}$, $M=M(\mu_{\rm X})$, which should be determined by the mean-field gap equation
\begin{eqnarray}
\frac{\partial \Omega_{\rm MF}(\mu_{\rm X},M)}{\partial M}=0.
\end{eqnarray}
The charge susceptibility can be evaluated as
\begin{eqnarray}
(\kappa_{\rm X})_{\rm MF}=-\frac{\partial^2\Omega_{\rm MF}(\mu_{\rm X},M)}{\partial \mu_{\rm X}^2}-\frac{\partial^2\Omega_{\rm MF}(\mu_{\rm X},M)}{\partial \mu_{\rm X}\partial M}
\frac{\partial M(\mu_{\rm X})}{\partial\mu_{\rm X}}.
\end{eqnarray}
The quantity $\partial M/\partial\mu_{\rm X}$ can be deduced from the gap equation. We have
\begin{eqnarray}
\frac{\partial^2\Omega_{\rm MF}(\mu_{\rm X},M)}{\partial \mu_{\rm X} \partial M }+\frac{\partial^2\Omega_{\rm MF}(\mu_{\rm X},M)}{\partial M^2}
\frac{\partial M(\mu_{\rm X})}{\partial\mu_{\rm X}}=0,
\end{eqnarray}
which leads to
\begin{eqnarray}
\frac{\partial M(\mu_{\rm X})}{\partial\mu_{\rm X}}=-\frac{\partial^2\Omega_{\rm MF}(\mu_{\rm X},M)}{\partial \mu_{\rm X} \partial M }\left[\frac{\partial^2\Omega_{\rm MF}(\mu_{\rm X},M)}{\partial M^2}\right]^{-1}.
\end{eqnarray}
Hence we obtain
\begin{eqnarray}
(\kappa_{\rm X})_{\rm MF}=-\frac{\partial^2\Omega_{\rm MF}(\mu_{\rm X},M)}{\partial \mu_{\rm X}^2}+\left[\frac{\partial^2\Omega_{\rm MF}(\mu_{\rm X},M)}{\partial \mu_{\rm X}\partial M}\right]^2
\left[\frac{\partial^2\Omega_{\rm MF}(\mu_{\rm X},M)}{\partial M^2}\right]^{-1}.
\end{eqnarray}
On the other hand, from the linear response theory, we have
\begin{eqnarray}
\Pi_{\rm MF}^{00}(Q)=\Pi_{\rm b}^{00}(Q)-\sum_{\rm m=0}^3\frac{C^0_{\rm
m}(Q)C^0_{\rm m}(-Q)}{\frac{1}{2G}+\Pi_{\rm m}(Q)}.
\end{eqnarray}
In the static and long-wavelength limit $Q\rightarrow0$, we have $C^0_{\rm m}(Q)\rightarrow0$ for ${\rm m}=1,2,3$. Thus we obtain 
\begin{eqnarray}
\Pi_{\rm MF}^{00}(Q\rightarrow0)=\Pi_{\rm b}^{00}(Q\rightarrow0)-\frac{\left[C^0_0(Q\rightarrow0)\right]^2}{\frac{1}{2G}+\Pi_{0}(Q\rightarrow0)}.
\end{eqnarray}
Using the explicit form of the above functions, we can show that
\begin{eqnarray}
\Pi_{\rm b}^{00}(Q\rightarrow0)&=&\frac{\partial^2\Omega_{\rm MF}(\mu_{\rm X},M)}{\partial \mu_{\rm X}^2},\nonumber\\
C^0_0(Q\rightarrow0)&=&\frac{\partial^2\Omega_{\rm MF}(\mu_{\rm X},M)}{\partial \mu_{\rm X} \partial M },\nonumber\\
\frac{1}{2G}+\Pi_{0}(Q\rightarrow0)&=&\frac{\partial^2\Omega_{\rm MF}(\mu_{\rm X},M)}{\partial M^2}.
\end{eqnarray}
Thus the compressibility sum rule is satisfied in the mean-field theory, i.e.,
\begin{eqnarray}
\Pi_{\rm MF}^{00}(Q\rightarrow0)=-(\kappa_{\rm X})_{\rm MF}.
\end{eqnarray}

When the meson fluctuations are taken into account, we have
\begin{eqnarray}
\kappa_{\rm X}=(\kappa_{\rm X})_{\rm MF}+(\kappa_{\rm X})_{\rm FL},
\end{eqnarray}
where the meson-fluctuation contribution can be evaluated as
\begin{eqnarray}
(\kappa_{\rm X})_{\rm FL}&=&-\frac{\partial^2\Omega_{\rm FL}(\mu_{\rm X},M)}{\partial \mu_{\rm X}^2}-2\frac{\partial^2\Omega_{\rm FL}(\mu_{\rm X},M)}{\partial \mu_{\rm X}\partial M}
\frac{\partial M(\mu_{\rm X})}{\partial\mu_{\rm X}}-\frac{\partial^2\Omega_{\rm FL}(\mu_{\rm X},M)}{\partial M^2}\left[\frac{\partial M(\mu_{\rm X})}{\partial\mu_{\rm X}}\right]^2\nonumber\\
&&-\frac{\partial\Omega_{\rm FL}(\mu_{\rm X},M)}{\partial M}\frac{\partial^2 M(\mu_{\rm X})}{\partial\mu_{\rm X}^2}.
\end{eqnarray}
We note that the effective quark mass $M(\mu_{\rm X})$ is still determined by the mean-field gap equation. On the other hand, the meson-fluctuation contribution to the density response function can be decomposed as
\begin{eqnarray}
\Pi_{\rm FL}^{00}(Q)=\Pi_{\rm OP}^{00}(Q)+\Pi_{\rm AL}^{00}(Q)+\Pi_{\rm SE}^{00}(Q)+\Pi_{\rm MT}^{00}(Q).
\end{eqnarray}
We can show that the first three terms in $(\kappa_{\rm X})_{\rm FL}$ is related to the sum of AL, SE, and MT contributions in the $Q\rightarrow0$ limit,
\begin{eqnarray}
&&\lim_{Q\rightarrow0}\left[\Pi_{\rm AL}^{00}(Q)+\Pi_{\rm SE}^{00}(Q)+\Pi_{\rm MT}^{00}(Q)\right]\nonumber\\
&=&\frac{\partial^2\Omega_{\rm FL}(\mu_{\rm X},M)}{\partial \mu_{\rm X}^2}+2\frac{\partial^2\Omega_{\rm FL}(\mu_{\rm X},M)}{\partial \mu_{\rm X}\partial M}
\frac{\partial M(\mu_{\rm X})}{\partial\mu_{\rm X}}+\frac{\partial^2\Omega_{\rm FL}(\mu_{\rm X},M)}{\partial M^2}\left[\frac{\partial M(\mu_{\rm X})}{\partial\mu_{\rm X}}\right]^2.
\end{eqnarray}
To prove this, we recall that in the presence of only $A_0(Q)$, the induced perturbations are given by
\begin{equation}
\eta_{\rm m}(Q)=-\frac{C^0_{\rm m}(-Q)A_0(Q)}{\frac{1}{2G}+\Pi_{\rm m}(Q)}+O(A^2),\
\ \ \ \ \ \eta_{\rm m}(-Q)=-\frac{C^0_{\rm m}(Q)A_0(-Q)}{\frac{1}{2G}+\Pi_{\rm
m}(Q)}+O(A^2).
\end{equation}
In the limit $Q\rightarrow0$, only $\eta_{0}$ survives and hence
\begin{equation}
\lim_{Q\rightarrow0}\frac{\eta_0(Q)}{A_0(Q)}=-\lim_{Q\rightarrow0}\frac{C^0_0(-Q)}{\frac{1}{2G}+\Pi_0(Q)}=\frac{\partial M(\mu_{\rm X})}{\partial\mu_{\rm X}}.
\end{equation}
The order parameter induced contribution,  $\Pi_{\rm OP}^{00}(Q)$, is related to the last term in $(\kappa_{\rm X})_{\rm FL}$. We have
\begin{eqnarray}
\Pi_{\rm OP}^{00}(Q\rightarrow0)={\cal C}_0{\cal R}_2^{00}(Q\rightarrow 0).
\end{eqnarray}
Using the fact that
\begin{equation}
{\cal C}_0=\frac{\partial\Omega_{\rm FL}(\mu_{\rm X},M)}{\partial M},  \ \ \ \ \ \ \lim_{Q\rightarrow0}{\cal R}_2^{00}(Q)=\frac{\partial^2 M(\mu_{\rm X})}{\partial \mu_{\rm X}^2},
\end{equation}
we find that the the last term in $(\kappa_{\rm X})_{\rm FL}$ is exactly given by the OP contribution. We can further understand this result by working out the explicit form 
\begin{eqnarray}
\frac{\partial^2 M(\mu_{\rm X})}{\partial \mu_{\rm X}^2}&=&-
\left\{\frac{\partial^3\Omega_{\rm MF}(\mu_{\rm X},M)}{\partial \mu_{\rm X}^2\partial M}+2\frac{\partial^3\Omega_{\rm MF}(\mu_{\rm X},M)}{\partial \mu_{\rm X}\partial M^2}
\frac{\partial M(\mu_{\rm X})}{\partial\mu_{\rm X}}+\frac{\partial^3\Omega_{\rm MF}(\mu_{\rm X},M)}{\partial M^3}\left[\frac{\partial M(\mu_{\rm X})}{\partial\mu_{\rm X}}\right]^2\right\}\nonumber\\
&&\times\left[\frac{\partial^2\Omega_{\rm MF}(\mu_{\rm X},M)}{\partial M^2}\right]^{-1}.
\end{eqnarray}
 
In summary, we have shown that the compressibility sum rule is exactly satisfied in the linear response theory including the meson fluctuations. The order parameter induced contribution is rather crucial to
recover the correct static and long-wavelength limit.

\subsection{The chiral symmetry restored phase}
One special case we are interested is the chiral symmetry restored phase ($T>T_c$) in the chiral limit ($m_0=0$). In this case, we have ${\cal C}_0=0$ and hence the order parameter induced
contribution vanishes. Also, we have $C_{\rm m}^\mu(Q)=0$, indicating that we do not need to consider the induced perturbations $\eta_{\rm m}(Q)$. In this case, the formalism becomes rather simple and we can 
identity various contributions diagrammatically.

In the chiral symmetry restored phase, the sigma meson and pions become degenerate. We have
\begin{eqnarray}
[{\bf D}(Q)]_{\rm mn}={\cal D}(Q)\delta_{\rm mn},
\end{eqnarray} 
where the propagator of the mesonic modes above $T_c$ is given by
\begin{eqnarray}
{\cal D}^{-1}(Q)
&=&\frac{1}{2G}+N_cN_f \int{d^3{\bf k}\over
(2\pi)^3}\Bigg[\left(\frac{1-f(E_{\bf k}^+)-f(E_{{\bf k}+{\bf q}}^-)}{iq_l-E_{\bf k}-E_{{\bf k}+{\bf q}}}-\frac{1-f(E_{\bf k}^-)-f(E_{{\bf k}+{\bf q}}^+)}{iq_l+E_{\bf k}+E_{{\bf k}+{\bf q}}}\right)
\left(1+\frac{{\bf k}\cdot ({\bf k+q})}{E_{\bf k} E_{{\bf k}+{\bf q}}}\right)\nonumber\\
&&+\left(\frac{f(E_{\bf k}^-)-f(E_{{\bf k}+{\bf q}}^-)}{iq_l+E_{\bf k}-E_{{\bf k}+{\bf q}}}-\frac{f(E_{\bf k}^+)-f(E_{{\bf k}+{\bf q}}^+)}{iq_l-E_{\bf k}+E_{{\bf k}+{\bf q}}}\right)
\left(1-\frac{{\bf k}\cdot ({\bf k+q})}{E_{\bf k} E_{{\bf k}+{\bf q}}}\right)\Bigg].
\end{eqnarray}
Here $E_{\bf k}=|{\bf k}|$ for $T>T_c$. Due to the degeneracy of the sigma meson and pions, various contributions to the linear response above $T_c$ become simple.

\subsubsection{Aslamazov-Lakin contribution}
\begin{figure}
\centering
\includegraphics[width=3.5in]{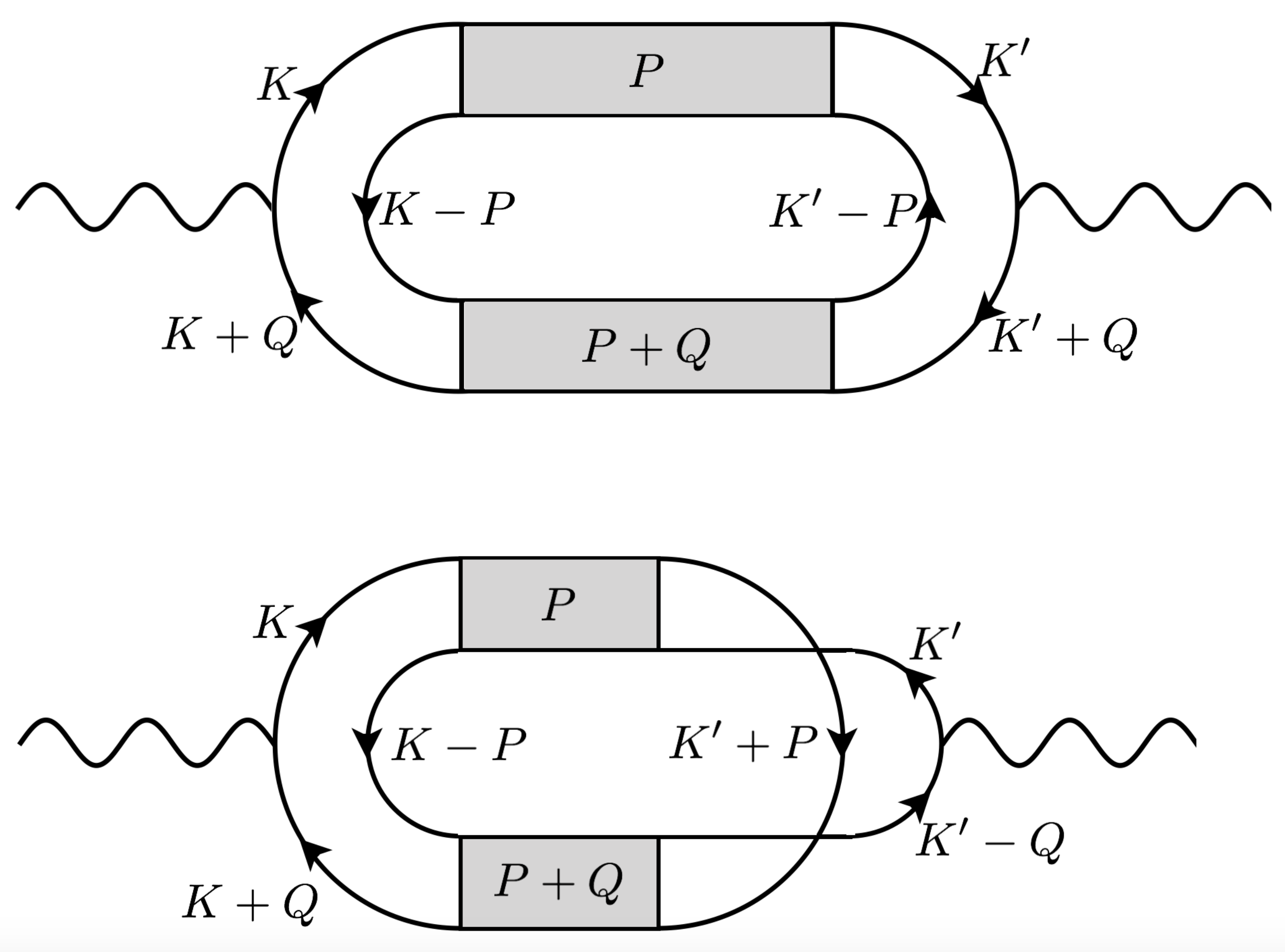}
\caption{Diagrammatic representation of the Aslamazov-Lakin contribution. Note that there are two kinds of AL-type diagrams. The solid lines with arrows denotes the quark propagator, the shaded boxes are the meson propagator,
and the wavy lines represents the external source.} \label{fig5}
\end{figure}

Above $T_c$, the Aslamazov-Lakin contribution  is given by
\begin{eqnarray}
{\cal W}_{\rm FL}^{({\rm AL})}=\frac{\beta V}{2}\sum_{Q} \Pi_{\rm AL}^{\mu\nu}(Q)A_{\mu}(-Q)A_{\nu}(Q),
\end{eqnarray}
where the AL response function $\Pi_{\rm AL}^{\mu\nu}(Q)$ is given by
\begin{eqnarray}
\Pi_{\rm AL}^{\mu\nu}(Q)=-\frac{2}{\beta V}\sum_{P}\left[{\cal D}(P){\cal D}(P+Q){\cal X}^\mu(P,Q){\cal X}^\nu(-P,-Q)\right].
\end{eqnarray}
The function ${\cal X}^\mu(P,Q)$ here is defined as
\begin{eqnarray}
{\cal X}^\mu(P,Q)&=&\frac{1}{\beta V}\sum_K{\rm Tr}
\left[{\cal G}(K)\Gamma^{\mu}{\cal G}(K+Q){\cal G}(K-P)\right]\nonumber\\
&+&\frac{1}{\beta V}\sum_K{\rm Tr}\left[{\cal G}(K-Q)\Gamma^{\mu}{\cal G}(K){\cal G}(K+P)\right].
\end{eqnarray}
Here the quark propagator ${\cal G}(K)$ is given in Eq. (85) with $M=0$. The AL contribution can be diagrammatically demonstrated in Fig. \ref{fig5}.

\subsubsection{Self-Energy contribution}
\begin{figure}
\centering
\includegraphics[width=3.5in]{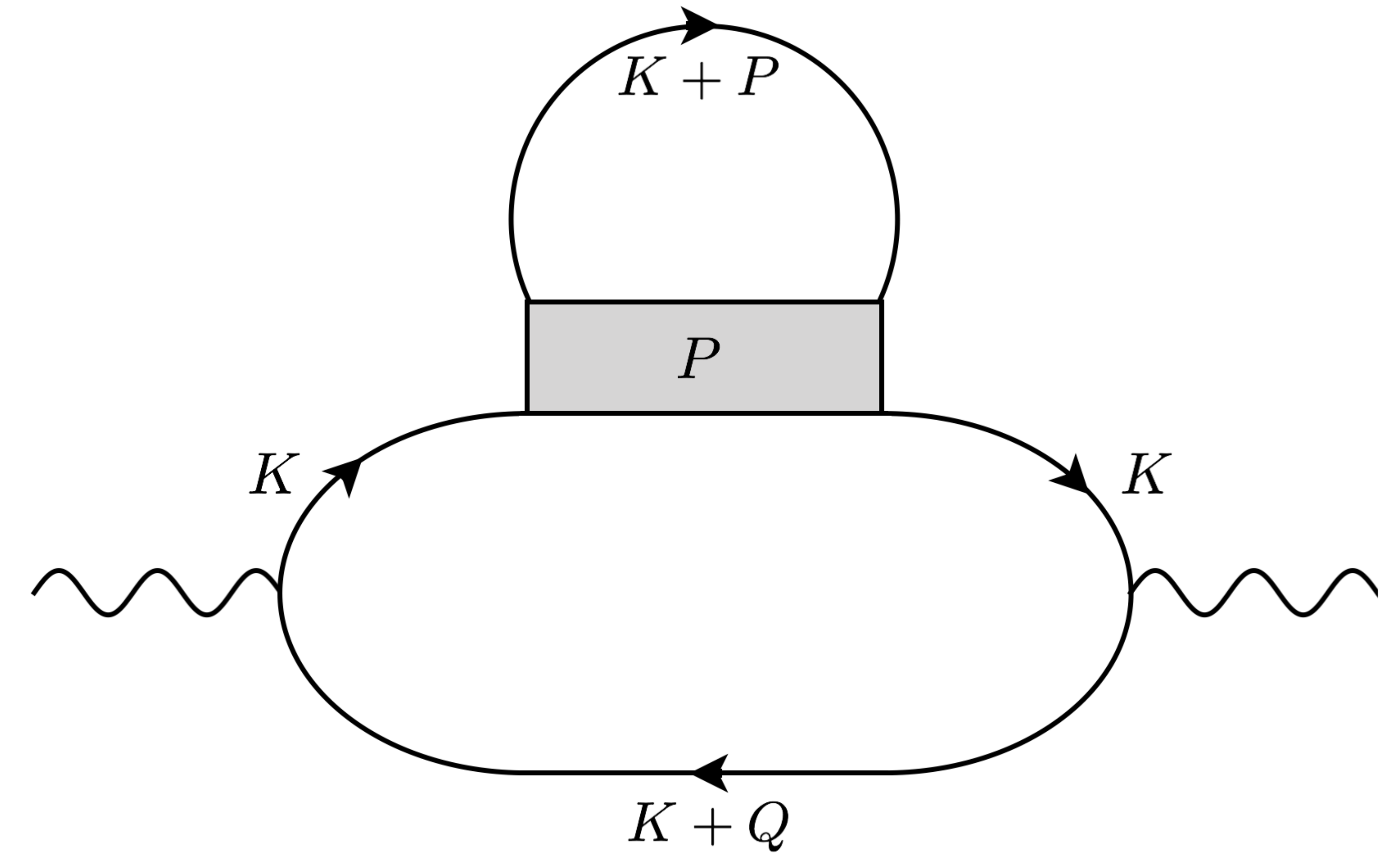}
\caption{Diagrammatic representation of the Self-Energy contribution. The notations are the same as in Fig. \ref{fig5}. } \label{fig6}
\end{figure}

Above $T_c$, the Self-Energy or Density-of-State contribution is given by
\begin{eqnarray}
{\cal W}_{\rm FL}^{({\rm SE})}=\frac{\beta V}{2}\sum_{Q} \Pi_{\rm SE}^{\mu\nu}(Q)A_{\mu}(-Q)A_{\nu}(Q),
\end{eqnarray}
where the SE response function $\Pi_{\rm SE}^{\mu\nu}(Q)$ is given by
\begin{eqnarray}
\Pi_{\rm SE}^{\mu\nu}(Q)=\frac{8}{\beta V}\sum_{P}\left[{\cal D}(P){\cal Y}^{\mu\nu}(P,Q)\right].
\end{eqnarray}
Here the function ${\cal Y}^{\mu\nu}(P,Q)$ is explicitly given by
\begin{eqnarray}
{\cal  Y}^{\mu\nu}(P,Q)=\frac{1}{\beta V}\sum_{K}{\rm Tr}\left[{\cal G}(K)\Gamma^{\mu}{\cal G}(K+Q)\Gamma^{\nu}{\cal G}(K){\cal G}(K+P)\right].
\end{eqnarray}
Note that $\Pi_{\rm SE}^{\mu\nu}(Q)$ can also be written as
\begin{eqnarray}
\Pi_{\rm SE}^{\mu\nu}(Q)=\frac{2}{\beta V}\sum_{K}{\rm Tr}\left[\Gamma^{\mu}{\cal G}(K+Q)\Gamma^{\nu}{\cal G}(K)\Sigma_{\rm q}(K){\cal G}(K)\right],
\end{eqnarray}
where $\Sigma_{\rm q}$ is the quark self-energy,
\begin{eqnarray}
\Sigma_{\rm q}(K)=\frac{4}{\beta V}\sum_{P}\left[{\cal D}(P){\cal G}(K+P)\right].
\end{eqnarray}
The SE contribution can be diagrammatically demonstrated in Fig. \ref{fig6}.

\subsubsection{Maki-Thompson contribution}
\begin{figure}
\centering
\includegraphics[width=3.5in]{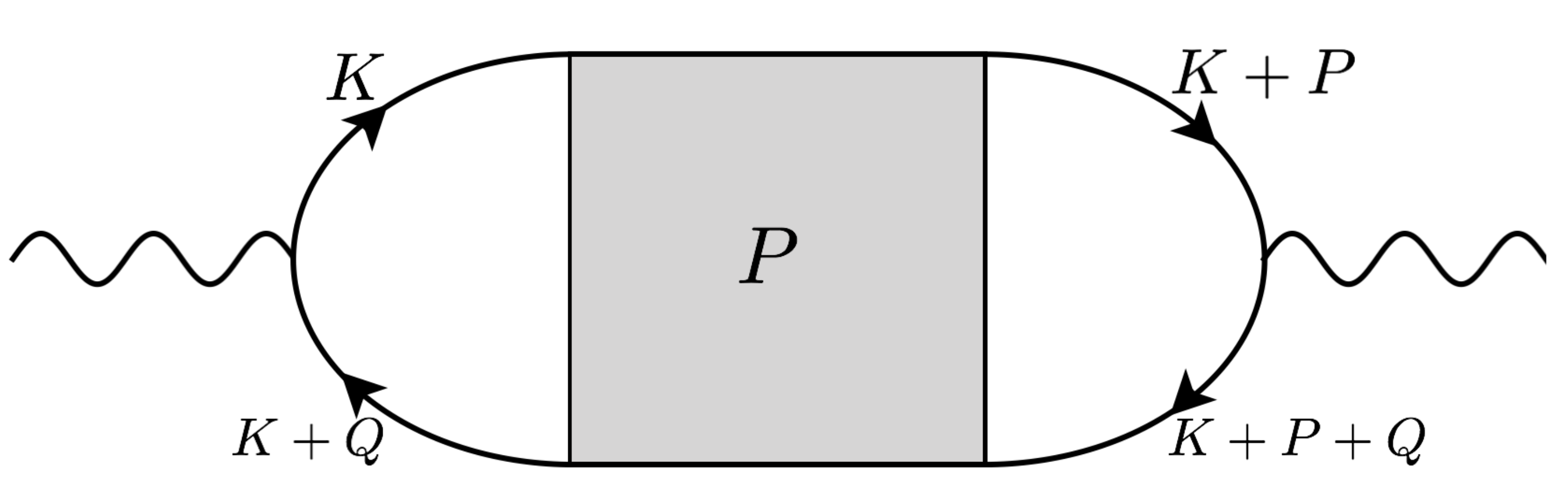}
\caption{Diagrammatic representation of the Maki-Thompson contribution. The notations are the same as in Fig. \ref{fig5}.} \label{fig7}
\end{figure}

Above $T_c$, the Maki-Thompson contribution is given by
\begin{eqnarray}
{\cal W}_{\rm FL}^{({\rm MT})}=\frac{\beta V}{2}\sum_{Q} \Pi_{\rm MT}^{\mu\nu}(Q)A_{\mu}(-Q)A_{\nu}(Q),
\end{eqnarray}
where the MT response function $\Pi_{\rm MT}^{\mu\nu}(Q)$ is given by
\begin{eqnarray}
\Pi_{\rm MT}^{\mu\nu}(Q)=\frac{4}{\beta V}\sum_{P}\left[{\cal D}(P){\cal W}^{\mu\nu}(P,Q)\right].
\end{eqnarray}
Here the function ${\cal W}^{\mu\nu}(P,Q)$ is explicitly given by
\begin{eqnarray}
{\cal W}^{\mu\nu}(P,Q)=\frac{1}{\beta V}\sum_{K}{\rm Tr}
\left[{\cal G}(K)\Gamma^{\mu}{\cal G}(K+Q){\cal G}(K+P+Q)\Gamma^{\nu}{\cal G}(K+P)\right].
\end{eqnarray}
The MT contribution can be diagrammatically demonstrated in Fig. \ref{fig7}.

In summary, the AL, SE, and MT contributions can be diagrammatically identified in the chiral symmetry restored phase. These contributions includes the propagator of the degenerate mesonic modes, ${\cal D}(Q)$.
Near the chiral phase transition temperature, these mesonic modes are soft modes and are nearly massless. Therefore, we expect that these mesonic modes may have significant effect on the transport properties  
near the transition. 

\section{Summary}\label{s8}

In this work, we have studied the linear response of hot and dense matter in the two-flavor Nambu-Jona-Lasino model. The linear response theory is formulated within the path integral approach. In this elegant formalism, 
the current-current correlation functions or the response functions can be conveniently calculated by introducing the conjugated external gauge field as an external source and expanding the generating functional in 
powers of the external source. Parallel to the well-established approximations for the equilibrium thermodynamics, we studied the linear response within the mean-field theory and a beyond-mean-field theory taking into account the mesonic contributions.

In the mean-field approximation, the response function recovers the quasiparticle random phase approximation. 
The dynamical structure factors for various density responses have been studied by using the random phase approximation. In the long-wavelength limit, the dynamical structure factors are nonzero only for the axial baryon density and the axial isospin density channels. 
For the axial isospin density channel, the dynamical structure factor can be used to reveal the Mott dissociation of pions at finite temperature. Below the Mott transition temperature, the dynamical structure factor reveals a pole plus continuum structure. Above the Mott transition temperature, it has only a continuum part. 

It is generally expected that the mesonic degrees of freedom are important both in the chiral symmetry broken and above and near the chiral phase transition temperature.  In the chiral symmetry restored phase, the random phase approximation describes the linear response of a hot and dense gas of noninteracting massless quarks.  Therefore, the mesonic degrees of freedom are not taken into account. In this work we have developed a linear response theory based on the meson-fluctuation theory which includes properly the mesonic degrees of freedom.  The mesonic fluctuations naturally give rise to three kinds of famous diagrammatic contributions: the Aslamazov-Lakin contribution, the Self-Energy or Density-of-State contribution, and the Maki-Thompson contribution. In the chiral symmetry breaking phase, we also found an additional chiral order parameter induced contribution, which ensures that the temporal component of the response functions in the static and long-wavelength limit recovers the correct charge susceptibility defined by using the equilibrium thermodynamic quantities. These contributions from the mesonic fluctuations are expected to have significant effects on the transport properties of hot and dense matter around the chiral phase transition or crossover, where the mesonic degrees of freedom are still important.

{\bf Acknowledgments:} The work is supported by the National Natural Science Foundation of China (Grant Nos. 11775123, 11890712, 11747312, and 11475062) and the National Key R\&D Program of China (Grant No. 2018YFA0306503).

\end{document}